\documentclass[aps, prd, twocolumn, lengthcheck, superscriptaddress, showpacs, letterpaper, nofootinbib]{revtex4-1}
\usepackage{amsmath,amssymb}
\usepackage{epsfig}
\usepackage{graphicx}
\usepackage{amsmath}
\usepackage{slashed}
\usepackage{amsfonts}
\usepackage{epstopdf}
\usepackage{color}
\newcommand{\be}{\begin{equation}}
\newcommand{\ee}{\end{equation}}

\newcommand{\ba}{\begin{array}}
\newcommand{\ea}{\end{array}}

\def\be{\begin{eqnarray}}
\def\ee{\end{eqnarray}}

\def\calB{{\cal B}}

\def\del{\partial}

\def\roughly#1{\mathrel{\raise.3ex\hbox{$#1$\kern-.75em%
\lower1ex\hbox{$\sim$}}}}
\def\lsim{\roughly<}
\def\gsim{\roughly>}

\def\la{{\Big<}}
\def\ra{{\Big>}}
\def\tr{\rm tr}
\def\bi{\bibitem}

\begin{document}

\title{Fractionalized Quasiparticles in Dense Baryonic Matter}
\author{Mannque Rho}
\email{mannque.rho@ipht.fr}
\affiliation{Universit\'e Paris-Saclay, CNRS, CEA, Institut de Physique Th\'eorique,  91191 Gif-sur-Yvette c\'edex, France }

\date{\today}

\begin{abstract}

I discuss a novel scenario on how baryonic matter could turn into a Fermi liquid of fractionalized baryons driven by high density inside compact stars. It involves the ``hidden" local vector bosons $\rho$ and $\omega$ together with the ``hidden" scale symmetry dilaton $\chi$, both  taken to be ``dual" to the quarks and gluons of QCD in the sense of quark $\leftrightarrow$ hadron duality  and offers a possibility, exploiting the ``Cheshire Cat Principle", to  unify the ``dichotomy" of  baryons described in terms of skyrmions and fractional quantum Hall (FQH) droplets {\it and} to explore how the baryonic matter could behave as density goes way beyond the normal nuclear matter density to the density regime where ``deconfined" quarks are presumed to figure. The question is raised as to whether the fractionalized characteristics of the ``stuff" -- whatever they are --  can be ``probed" in the dense star core of compact stars.

\end{abstract}
\maketitle
\setcounter{footnote}{0}

\section{Introduction}
A nucleon, proton (neutron), in nuclei  is thought in QCD, the fundamental theory of strong interactions, to consist of three confined quarks, 2 up (down) and one down (up) quarks. Each quark then carries 1/3 of  the baryon charge (number). It has not  been ``seen"  directly in experiments and there is no -- theoretically -- known way to reveal the 1/3-charged  baryon.  But we know from the celebrated Adler-Bell-Jackiw triangle anomaly via $\pi^0\to 2\gamma$ decay that the quark baryon charge should be $1/N_c$ where $N_c=3$ is the number of colors.  

The question one can legitimately ask is this: When embedded strongly correlated  in the system of large number of baryons, does the individual quark preserve its 1/3 baryon charge? For low enough density, at least up to, say, a nuclear matter density  $n_0\simeq 0.16$ fm$^{-3}$, nucleons in medium can be well described with baryons, each of which carrying  the baryon number 1. Thus it makes a great sense to treat  the lead-208 nucleus $^{208}$Pb in terms of 208 nucleons interacting via local nuclear interactions between them, not in terms of 208  bags such as in the MIT bag model each containing 3 confined quarks. In the Folk Theorem \`a la Weinberg~\cite{FT}, QCD in nuclear physics should be fully captured by a chiral effective field theory,  $\chi$EFT, at least at low energy and density, in terms of nucleons and pions instead of quarks and gluons.

But what happens when the system of nucleons is squeezed to high density, say, 5 to 7 times the density of nuclear matter ($n_0\simeq 0.16$ fm$^{-3}$) as one expects in massive compact stars or CBM experiments at GSI and elsewhere? 

As the confinement ``bags" -- if taken seriously -- start to overlap, the quarks inside the bags could most plausibly start to ``percolate"~\cite{percolation} between bags. There can be  various scenarios describing the percolation phenomena with or without phase transitions~\cite{baymetal}. The question would then be what are the ``stuffs" that percolate? The answer, more or less standard, could be that the stuffs are a bunch of quarks with baryon charge 1/3.  But in order to accommodate what's observed in nature, say, in compact stars, the interactions between the percolating quarks need to be considerably strengthened from those inside the bags~\cite{baymetal}.  

In the paucity of approaches that are model-independent with lattice QCD calculations unavailable at high densities, it is totally unclear how to address dense baryonic matter pertinent to massive compact stars.  Among the large number of model approaches, typically energy density functionals and effective field theories, the one closest to the fundamental theory of strong interactions QCD is an approach that resorts to the notion that the quarks are the indispensable degrees of freedom and invoke a series of sequential QCD phase transitions -- after the initial cross-over from nuclear matter to quark matter which is fraught with uncertainty -- all the way to the color-flavor-locked phase, as discussed, e.g.,  in \cite{alford}. 

Now given  strong correlations involved in the process, one wonders, why should one stick to the notion that the relevant degrees of freedom are the quarks and no other ``stuffs"?

{\it The principal question addressed in this note: What could be the ``stuff" in the core of massive compact stars? If the stuff is not a bunch of deconfined quarks, how can one relate that stuff to the ingredients of QCD?}

In condensed matter physics, there are a plethora of fractionalized electrons in highly correlated systems~\cite{tong}. Just to cite among many, the celebrated  example is the Laughlin's fractional quantized Hall state with $\nu=1/3$, where the electron carries charge  $e/3$, and in the case of half-filled lowest Landau level (LLL),  $\nu=1/2$ with the electron charge  $e/2$.  Furthermore the state of the matter is drastically different depending on whether the electron is $e/2$-charged or $e/3$-charged.

In this note, I would like to describe a possible scenario that in the highly compressed state of baryonic matter before reaching asymptotic density where the Collins-Perry deconfinement sets in~\cite{collins-perry}, the relevant degrees of freedom are ``quasiparticles" with fractional baryon charges. I suggest  to interpret  the changeover from the baryons familiar at low density to fractionally charged quasipartlcles at high density, relevant to, say, the cores of massive neutron stars, as what's termed the ``hadron-quark continuity" much discussed in the literature~\cite{baymetal}.  Later I will argue that ``hadron-quark continuity' implying the absence of ``quark deconfinement" is a misnomer. What is more appropriate is hadron-quark duality with the accent on ``duality." The terminology ``quasiparticles" is here used in place of either hadrons  or  quarks since  they could equally well represent either one or the other  in the sense of hadron-quark duality.   In formulating the basic thesis, I draw an analogy or perhaps a conjecture between the half-skyrmions in dense baryonic medium and  the recently proposed structure of the single-flavor baryon -- that will be referred to as $B^{(0)}$ -- as a  fractional quantum Hall droplet,  a pancake~\cite{zohar}  or, perhaps more pertinently a ``pita"~\cite{karasik}.  The issue that arises is at what density the notion of hadron-quark ``continuity" -- as opposed to ``duality" --  ceases to be relevant. I will argue that this issue would arise at  a density much higher than relevant to massive compact stars stable against gravitational collapse. 

Although the basic idea is quite simple, it is in some sense highly unorthodox and the details are involved. This possible ``defect" may be due to the insufficient understanding of some of the concepts involved. It could be removed as more data become available from up-coming experiments and incorrect ideas are weeded out. Up to date,  there seems to be nothing unmistakably or convincingly at odds with nature. My attitude here is that given that the approach that I advocate is {\it simple and verifiable} by future observations to come, I push it full-speed ahead until ``torpedoed" by rigorous theory or by experiments.

As interesting cases in possible support of the key idea, i.e., hadron-quark duality, I will cite two examples which as far as I am aware have not been addressed by other workers in the field. One is the long-standing puzzle of the ``quenched $g_A$" at low density in nuclear Gamow-Teller transitions which, it is argued, is connected to the ``dilaton-limit fixed point" at a much higher density~\cite{quenchedgA}.   And the other is a possible precocious onset of the conformal sound velocity $v_s^2/c^2=1/3$  in the interior of massive stars at a density a few times normal nuclear matter density, interpreted as a precursor to (approximate) scale invariance emerging from strong nuclear interactions.  

\section{Hadron-Quark Duality}
Strong interactions are lot more complicated, involving several different flavors than the electromagnetic interactions that govern condensed matter dynamics. So one cannot develop  simple and elegant descriptions closely aided by experiments often found in condensed matter physics. Furthermore accessing strong-interaction phenomena with experiments is very hard and indirect if at all feasible. The situation is much more challenging at high density since the latter, unlike at high temperature, cannot be accessed directly by QCD.  

At very high energy -- and also at asymptotic density, QCD tells us it is the quarks and gluons that are relevant actors. There are ample, firm,  indications from experiments that at high energy it is indeed so. But up to today there is no clear indication that that is the case at high density,  surpassing  that of normal nuclear matter. There are two reasons for this situation. One, QCD, as is by now generally accepted,  cannot access high density in a reliable way as no workable theories nor models are available for guidance. Two, in the absence of trustful theoretical tools, there are no clear experimental ways to unravel the complex phenomena to pinpoint the relevant degrees of freedom. This specially applies to what's happening in physics of dense compact stars. There is a wealth of papers detailing compact-star data and model analyses of terrestrial experiments.  Some of the results published in the literature can explain more or less -- and some exceptionally well -- {\it all} the experimentally measured data, but this feat is achieved at the cost of large number of adjustable parameters uncontrolled by rigorous theory.  The problem with this ``successful" fitting is that it teaches us nothing of what actually is taking place.  Whatever is something ``new" so far has been explained by certain models by fiddling parameters still at one's disposal, which implies that no new physics, if there is any, is uncovered.  

So the question is ``what are the smoking-guns observable that could weed out wrong theories at high density?"

To address this issue, let me start by first assuming the basic properties of QCD that quarks are confined at nonasymptotic densities and obey (nearly perfect) chiral symmetry. At low density, say, in nuclei and nuclear matter, the relevant degrees of freedom are undoubtedly nucleons and low-mass mesons, i.e., pions as we know from nuclear effective field theories anchored on chiral symmetry that can well describe them at low energy. See  \cite{van kolck} for an up-to-date status . But what happens as density goes way up beyond that of normal nuclear matter? As mentioned, even if we have no clear theory of what will happen, it is very likely that when the bags overlap, quarks ``confined" within a bag start percolating into neighboring  bags. I take this a working hypothesis.

Now one of the ways to ```see" how  quarks transform to  hadrons and vice versa can be best illustrated by the Cheshire Cat phenomenon put forward a long time ago~\cite{CCP}. This   will be the strategy adopted in this note.

\subsection{``Infinite Hotel" for $N_f=2$: Skyrmions}

What takes place can be imagined as a quark in a jail trying to escape from the jail fully occupied like the filled Dirac sea\footnote{This is beautifully described in \cite{jail}. Actually $N_c$ quarks are involved but we focus on only one of them.}.  A massless quark swimming on top of the sea, say,  to the right  in one spatial dimension\footnote{The argument can  be straightforwardly extended to 3 spatial dimensions.}  in an attempt to escape the jail  bumps into the ``jail wall," so is unable  to escape.  It cannot swim back on top of the Dirac sea,  because chiral symmetry forbids it. But it can plunge into the Dirac sea which is feasible because the Dirac sea is infinite, a story of regularization, and swim back inside (to the left). This infinite Dirac sea can be likened to an  ``infinite hotel (IH)"~\cite{jail}.  This exploitation of the infinity  is  a quantum effect known as ``anomaly" in gauge theory.

There is one serious problem in this scenario, however. The fermion (baryon) charge carried by the quark disappears into the Dirac sea, so  the baryon number is apparently ``violated" in the process.  In QCD, the baryon charge is absolutely conserved, so the fermion charge cannot disappear. Here takes place a miracle.  The fermion charge is relayed to the ``pion"  that clouds the outside wall,  the pion (boson) turning into a baryon (fermion). This is by now the well-known story of skyrmions mathematically characterized by the homotopy group $\pi_3 (S^3)=\mathcal{\cal Z}$ for the $N_f\geq 2$ systems.

This IH phenomenon can be considered to involve two domains, one the quark-gluon one and the other the hadronic one. There are two modes of a global symmetry, i.e., chiral symmetry, involved:  Wigner-Weyl (WW) mode inside the bag and Nambu-Goldstone (NG) mode outside the bag. Therefore the jail wall can be taken as a ``domain wall (DW)" that delineates two vacua. This is the ``jail-break" scenario for the $N_f=2$ (i.e., proton and neutron) case. 

The upshot is that the leaking baryon charge is taken up by the pion as a soliton. So in nuclear physics, we argue that for the given soliton chiral angle  $\theta(R)$, the leaking baryon charge $1-\theta (R)/\pi$ (in 1 spatial dimension) is lodged in the skyrmion cloud while the rest of the  charge  $\theta (R)/\pi$ remains in the bag, yielding the total baryon number 1 for a single baryon.  When  the bag is infinite the whole baryon charge is lodged  inside the bag, while when the bag shrinks to zero size the whole baryon charge goes into the skyrmion cloud. The confinement size $R$ is therefore an unphysical quantity. One can think of this process as the pion fields giving rise to the baryons as solitons. This is what is referred to as ``Cheshire Cat Phenomenon" or ``Cheshire Cat Principle (CCP)"~\cite{CCP} This   is akin to the disappearance of the Cheshire Cat in ``Alice in the Wonderland" with baryon number playing the role of the cat's smile.  In fact it is perhaps more appropriate to identify this phenomenon as a gauge artifact and formulate a gauge theory for the phenomenon~\cite{CC-gauge}.

This discussion of the CC ``smile" applies straightforwardly to (3+1) dimensions. It has indeed been verified by Goldstone and Jaffe~\cite{GJ}  in terms of the spectral asymmetry $\eta(s)=\sum_n\epsilon(E_n)|E_{n}|^{-s}$ which gives the baryon charge lodged inside the bag for a given chiral angle $\theta (R)$. The fractionalization of the baryon charge is exact thanks to the topology involved. In (1+1) D, an exact bosonization allows an in-principle CCP for non-topological processes also,   but in the absence of bosonization in (3+1)d, exact CCP does not exist in the nuclear processes that we are concerned with, so much of what I will discuss below for processes that are not topological is at best approximate.

\subsection{Fractional quantum Hall droplet or pita}
The IH scenario discussed above famously does not work in 3d when the number of flavors is one. This is because $\pi_3 (U(1))=0$. One then wonders whether there is no soliton for baryon coming from the flavor singlet meson $\eta^\prime$. This puzzle was recently resolved by ideas developed in  condensed matter physics. It has been suggested by Komargodski~\cite{zohar} that the $\eta^\prime$ can ``turn into" a flavor singlet baryon denoted from here on as $B^{(0)}$ as a fractional quantum Hall (FQH) droplet. At first sight this  FQH droplet (pancake or pita) is unrelated to the usual skyrmion corresponding to the nucleon. 

Is there any relation between the two topological objects, the FQH droplet for $B^{(0)}$ and the skyrmion for nucleons? This is the question raised below -- a ``dichotomy" -- in connection with compact-star physics.
\subsubsection{Baryon for $N_f=1$}
It has been shown that  Komargodski's FQH pancake model can be given a simple formulation in terms of a Cheshire Cat phenomenon~\cite{MNRZ}. 

Suppose the quark in the bag is of $N_f=1$ in the jailbreak scenario.  Let the quark be coupled at the wall $x=R$ to the flavor-singlet meson $\eta^\prime$ assuming spontaneously broken chiral symmetry. We take the large $N_c$ limit. Again the confinement leads to the breaking of the baryon charge by the bag boundary condition and gives rise to the anomaly as in the $N_f\neq 1$ case, but since $\pi_3 (U(1))=0$, the quark cannot go into the infinite hotel. So where does it go?
The answer~\cite{MNRZ} is that the quark moving in, say,  the $x$ direction  flows in the $y$ direction into a  2d quantum Hall-type pancake, taking care of the anomaly and keeping the baryon charge conserved. This has been interpreted in \cite{MNRZ} as the ``anomaly in-flow" mechanism  leading to the Chern-Simons topological term ~\cite{callan-harvey},  which in 3-form reads 
\be
\frac{N_c}{4\pi}\int_{2+1} ada\label{cs}
\ee
where  $a_\mu$  is the Chern-Simons field that encodes strong correlations in QCD. When the baryon charge leaks completely into the FQH ``pancake," the ``smile" reduces to a vortex line on the pancake. {\it This suggests extending to the $N_f=1$ baryon the CCP that the ``confinement size" has no physical meaning as in the case of $N_f\neq1$ case~\cite{CCP}, sharpening the notion of hadron-quark duality.}  That the resultant FQH droplet correctly carries the baryon charge  is assured by the gauge invariance of the Chern-Simons term (\ref{cs}). This  can be explained in terms of a chiral bosonic edge mode ~\cite{karasik} which has been  identified with the $U(1)$ gauge field in hidden local symmetry (HLS). We will argue  this identification plays a possibly crucial role in accessing the EoS for massive compact stars in the approach detailed in \cite{MR-review,PKLMR}. 
  
Now in accordance with the global symmetries of QCD,   the $B=1$ baryon with $N_c$ quarks must then have spin $J=N_c/2$. This yields the high-spin baryon. Thus when the bag is shrunk to zero size, the Cheshire Cat smile resides in the vortex line in the FQH droplet. For instance for $N_c=3$, this picture yields the $\Delta (3/2,3/2)$.  The same  $\Delta (3/2,3/2)$ also appears in the rotational quantization of the skyrmion with $N_f=2$ which comes from the $\infty$-hotel mechanism. These two descriptions present an aspect of the {\it ``dichotomy problem"} mentioned above. The question is: Whether or how they are related?
\subsubsection{Pancake baryons for  $N_f=2$?}\label{dp}
Instead of a flavor-singlet quark, now consider the jail-breaking scenario of the doublet  u  and  d quarks. There seems to be  nothing to forbid the quark from flowing, instead of dropping into the infinite hotel, into the $y$ direction as the flavor-singlet quark did to compensate the anomaly generated by the bag wall.  Suppose one applies the same anomaly-flow argument in  CCP to the $N_f$-flavored quark. The spin-flavor symmetry for the flavor $N_f\neq 1$ will of course be different. Given $N_f=2 $ for instnace,  we expect to have a non-abelian Chern-Simons field $\mathbb A_\mu$ in place of the abelian $a_\mu$~\cite{MNRZ},
\be
\frac {N_c}{4\pi}\int_{2+1} {\rm Tr}\left(\mathbb A d\mathbb A+\frac 23 \mathbb A^3\right).
\label{13}
\ee
This presents an alternative jail-break scenario to the infinite-hotel one. Then the question: What directs the $N_f=2$ quark to go to either of the two possibilities, (A) drop into the infinite hotel or (B) flow into the FQH droplet?  Or (C) could it go to the infinite hotel at some density and flow as density increases to the FQH pancake (or pita) at some higher density? This possibility seems plausible.  A scenario along this line will be discussed in Section \ref{CCP2densematter}.

%
\subsection{Fermion number and Hall conductivity\\ on domain wall}\label{domainwall}
It has been argued~\cite{vassi} that the same Chern-Simons  structure can be arrived at with the spectral asymmetry $\eta$ employed in \cite{GJ} in (3+1)d with a domain wall (or an interface). For this we consider quantized Dirac fermions in interaction with a background $U(1)$ gauge field $a_\mu$, and scalar $\sigma$ and pseudo-scalar $\pi$ fields
\be
\cal{L}=\bar{\psi}{\cal  D}\psi
\ee
with 
\be
{\cal D}=i\gamma^\mu (\del_\mu-i g a_\mu)-(\sigma+i\gamma_5 \pi), \ \sigma^2 +\pi^2=1.
\ee
What the $U(1)$  gauge field $a_\mu$ and the gauge coupling $g$ represent are specified below in making  connection with the Cheshire Cat phenomenon. 
The vacuum baryon number $B$ is given by 
\be 
B=-\frac 12\eta(0,H)
\ee
where $\eta (s, H)$ is the  spectral asymmetry that was computed in \cite{GJ} (for the infinite --hotel scenario)
\be
\eta (s, H)=\sum_{\lambda >0} \lambda^{-s} -\sum_{\lambda<0} (-\lambda)^{-s}
\ee
where $\lambda$ is the eigenvalues of the Dirac Hamiltonian $H$.  
With some reasonable approximations, it was obtained in \cite{vassi} with a domain wall at $x_3=0$  that
\be
B=-\frac{g}{4\pi^2} \theta|^{x^3=+\infty}_{x^3=-\infty}\int d^2x f_{12}
\ee
where $\theta = ({\rm arctan}(\pi/\sigma))$ -- that we will identify with the ``chiral angle" later -- and $f_{\mu\nu}$ is the gauge field tensor. Note that the vacuum fermion number $B$ has two components, first the Goldstone-Wilczek fractionalized fermion number~\cite{goldstone-wilczek} and the other the magnetic flux through the $(x,y)$ 2-d plane  

Consider next a domain background defined by the fields $\sigma$ and $\pi$  that depend on $x^3$ only. The one-loop effective action in the non-static background, in (3+1)d gives the parity-odd action
\be
S=\int d^4x d^4y G(x,y)a_\mu (x)\del_\nu^y a_\rho (y) \epsilon^{\mu\nu\rho 3} 
\ee
where $G$ is a complicated non-local function of $x^3$, $y^3$ and $z^\alpha=x^\alpha-y^\alpha$, $\alpha=0,1,2$. In the long-wavelength limit in the form factor $G$,
the action can be written as a Chern-Simons term 
\be
S=g^2 \frac{k}{4\pi}\int d^3y^\alpha a_\mu(y^\alpha,0)\del_\nu a_\rho (y^\alpha,0)\epsilon^{\mu\nu\rho 3}\label{S}
\ee
with
\be
g^2\frac{k}{4\pi} = \int d^3x^\alpha dy^3 dx^3 G(z^\alpha, x^3,y^3).\label{k}
\ee
Here $k$ can be identified as the ``level" in the level-rank duality of the Chern-Simons term. 

To make contact with what was done in the CCP structure~\cite{MNRZ} we go to the domain wall structure corresponding to  the Cheshire Cat.  For this,  one may take the $U(1)$ field  to be the $\omega$ field when the vector mesons $\rho$ and $\omega$ in hidden local symmetry are treated as the color-flavor locked $U(N_f)$ gauge fields dual to the gluon fields in QCD~\cite{KKYY}. Then the  $\omega$ field can be taken as the Chern-Simons field that captures \`a la CCP the strongly-correlated excitations inside the region ``modeled" by the bag.\footnote{``Strongly-correlated excitations `modeled' by the bag" do not necessarily represent the MIT bag with the bag boundary conditions etc. They stand more generally for the confinement and interactions associated with the generic confinement mechanism.  }  Now for $U(N_f)_{-N_c}$ dual to $SU(N_c)_{N_f}$ spontaneously broken, the vortex configurations in three dimensions made up of $\rho$ and $\omega$ carry magnetic and electric charges of $U(1)^{N_f}$. The electric charge in the CS term can then be identified with  the baryon charge~\cite{KKYY}. This allows one to obtain  the vector current from the action $S$ (\ref{S}), the time component of which is
\be
J^0 (x)=\frac{1}{g}\frac{\delta}{\delta a_0(x)} S.
\ee
The baryon number is~\cite{vassi} 
\be
B=\int d^3x J_0 (x)=\frac{gk}{2\pi}\int f_{12} d^2x.
\ee
Setting the Dirac quantization for the magnetic flux threading the vortex~\cite{KKYY}
\be
\frac{g}{2\pi}\int  f_{12} d^2x =1
\ee
we will have that
\be 
B=k.
\ee 

Now to make the connection~\cite{vassi} to the Cheshire Cat scenario discussed in \cite{MNRZ}, we identify $\theta=\eta^\prime/f_{\eta^\prime}$ and impose at $x^3=0$ the Cheshire Cat boundary condition 
\be
(1-i\gamma^3 e^{i\gamma_5\theta})\psi|_{x^3=0}=0. \label{bc}
\ee
Then we obtain the baryon charge residing inside the bag fractionalized to
\be
\Delta B=\frac{\Delta \theta}{2\pi}
\ee
where  $\Delta\theta$ is the jump of the $\eta^\prime$ field across the chiral bag. This is the same result obtained in \cite{MNRZ}. 

Here are two crucial points, among others, to note. First of all, as pointed out in \cite{vassi},  the Chern-Simons term (\ref{S}) is not topological. This is because $k$ as defined in (\ref{k}) is not in general quantized, which means that the Lagrangian is not gauge invariant. The total baryon is of course conserved, so the total must restore the gauge invariance implying the anomaly cancellation. This must be related to the color anomaly discussed in \cite{colorleakage}. Second one could have done the same analysis for the $N_f=2$ case with the pion field giving rise to nonabelian CS theory with the same results as in the CCP strategy.

\section{Dichotomy problem}

The alternative jail-break scenarios of Section \ref{dp} ((A), (B) or (C)) present the dichotomy problem to be resolved: What makes the $N_f$-flavored  quarks go into either the IH or the FQH droplet or both while conserving the baryon charge? 
\subsection{Hidden symmetries}
A possible solution to this question is suggested by Karasik~\cite{karasik,karasik2}\footnote{In \cite{KM}, Kitano and Matsudo present an argument that seems to be in disagreement with Karasik's on how HLS figures in ``duality" with QCD. However both agree that HLS must play a crucial role at, say, chiral and/or deconfinement transition. I have not yet analyzed how the KM scenario fits in what's discussed in this note.} to involve hidden local symmetry (HLS)~\cite{HY:PR} for the low-lying vector mesons $\rho$ and $\omega$, though implicitly for our treatment, scale symmetry for scalar dilaton $\chi$. I am proposing that this can lead to a new development in nuclear physics, since it has been observed~\cite{dichotomy}  that the same scale-symmetric hidden local symmetry (``sHLS" for short) exploited in \cite{karasik} could play a crucial, hitherto-unexplored, role in massive compact-star physics~\cite{MR-review,PKLMR}.  The nature of hidden symmetries appropriate for the matter in discussion is spelled out generally in \cite{MR-review} and much more relevantly to the specific issue treated in this note in \cite{dichotomy}.

The argument made by Karasik~\cite{karasik} to combine $N_f=1$ and $N_f\geq 2$ in one unified form relies on  the {\it unified} baryon current 
\be
\calB_{unif}^\mu &=& \frac{1}{24\pi^2} \epsilon^{\mu\nu\rho\sigma} {\tr}\Big[2\del_\nu\xi\xi^\dagger\del_\rho\xi\xi^\dagger \nonumber\\
&+& 3i V_\nu (\del_\rho\xi\del_\sigma\xi^\dagger-\del_\rho\xi^\dagger\del_\sigma\xi)\nonumber\\
&+& 3i\del_\nu V_\rho (\del_\sigma\xi\xi^\dagger -\del_\sigma\xi^\dagger\xi) \Big]\label{gB}
\ee
where (in unitary gauge in HLS)
\be
U=\xi^2=e^{i\eta^\prime/2} (\sigma+i\pi_a \tau_a),\ \sigma^2+\pi_a^2=1\label{U}
\ee
and
\be
V_\mu=\frac 12 (\omega_\mu +\tau_a \rho_\mu^a).
\ee
The current (\ref{gB}), applied to {\it smooth} configurations,  gives the usual skyrmion baryon current  for $N_f\geq 2$ baryons 
\be
\calB^\mu_{N_f\geq 2}= \frac{1}{24\pi^2} \epsilon^{\mu\nu\rho\sigma} {\tr}&\Big[&2\xi^\dagger\del_\nu\xi\xi^\dagger\del_\rho\xi\xi^\dagger\del_\sigma\xi\nonumber\\
&-& 3\del_\nu\xi\del_\rho\xi\del_\sigma(\xi^\dagger)^2\Big].
\ee
What's new and significant is that it also accounts for {\it non-smooth} configurations such as FQH droplet with $N_f=1$ for which the baryon current  is\footnote{Note that the difference between ${\calB}_{unif}^\mu$ and ${\calB}^\mu_{N_f\geq 2}$ is a total derivative and hence there is a boundary term for a compact space, i.e., the pancake topology.} 
\be
{\calB}_{N_f=1}= -\frac{1}{8\pi^2}\epsilon^{\mu\nu\rho\sigma} \del_\nu\omega_\rho\del_\sigma \eta^\prime.\label{pancakeB}
\ee
This current comes from the homogeneous Wess-Zumino (hWZ) term in hidden local symmetry Lagrangian~\cite{HY:PR}\footnote{The coefficient $c_3$ in (3.163) of \cite{HY:PR}, which is an arbitrary coefficient for $N_f\geq 2$ associated with the ``dynamical gauge fields"  is set $c_3\to 1$ in arriving at the $B^{(0)}$. This may be constrained in deforming  ${\calB}_{unif}$ to ${\calB}_{N_f=1}$. This comes also from the vector dominance  in HLS theory~\cite{karasik2} where ``h" in hWZ stands for ``hidden" in place of ``homogeneous." This difference could be crucial for the physics involved.} which is a very important ingredient for dense baryonic matter treated as skyrmion matter with HLS Lagrangian~\cite{hwz-omega}.

The  dilaton field $\chi$ does not figure in the unified current (\ref{gB}), hence is  absent in (\ref{pancakeB}).  But  $\chi$ plays an indispensable role in the sHLS theory  in ``deforming" (\ref{gB}) to the flavor singlet baryon current (\ref{pancakeB}).  See also \cite{hwz-omega} for dense skyrmion matter. The $\omega$ field corresponds to  the Chern-Simons field $a_\mu$ in (2+1)D. To capture the Chern-Simons structure of the $B^{(0)}$, it is found necessary to have the condensate $\langle\chi\rangle\to 0$ and consequently the $\omega$ mass go to zero~\cite{dichotomy}\footnote{This scenario of driving the $\omega$ mass  to zero being done here by the dilaton condensate going to zero differs from the scenario in \cite{karasik2}  where the $\omega$ mass goes to zero as the quark mass is sent to infinity in the $\eta^\prime=\pi$ domain wall in discussing the duality of the $\omega$ to a glueball. This comes about because the quark-mass going to infinity on the $\theta=\pi$ domain wall makes the theory go to pure Y-M theory. How to reconcile the dichotomy of the $\omega$ as one approaches from bottom-up as is done in our scheme and the one coming top-down as in  \cite{karasik2} is not clear.}.
This is to assure that the baryon charge is correctly given with the Dirac condition 
\be
\frac{1}{2\pi}\int d\phi \omega_\phi\in {\cal Z}.
\ee
\subsection{The dichotomy}
The baryon current (\ref{gB}) is to encompass  baryons from $N_f=1$ to $N_f\geq 2$ and from light-quark baryons to heavy-liqht-quark baryons. An illustrative case is the baryon $\Delta (3/2.3/2)$.  As mentioned, it appears in the rotational quantization of the skyrmion (for $N_f\geq 2)$) and  in the FQH droplet (for $N_f=1$). {\it Naively}, the mass difference $\Delta M=m_\Delta-m_N$ is $\propto O(1/N_c) +\cdots$ in the former,  whereas in the latter it would be $\Delta M\propto O(N_c^0) +O(1/N_c)+\cdots$~\cite{zohar}. This $O(N_c^0)$ term present  in the FQH droplet but missing in the skyrmions is a signal for  the dichotomy problem~\cite{zohar,MNRZ}. 

Closely associated with this problem is  whether the flavor singlet baryon is  a  bona-fide observable quantity. If it is, how it can be probed?  

In \cite{karasik}, the dichotomy problem is posed as how the skyrmion  can be deformed to the FQH droplet and that by dialing the quark masses. 

The argument goes as follows. Consider the mass term in the Lagrangian
\be
{\cal L}_{mass}={\tr} (MU + h.c. -2M)
\ee
with $U$ given by (\ref{U}) and the mass matrix $M$ is  taken as
\be
M=
\Big (
\begin{array}{cc}
m_a\approx 0 & 0\\
0 & m_d
\end{array}
\Big).\\
\nonumber
\ee
In the large $N_c$ limit and for $m_d\approx  0$, the baryon is the usual skyrmion. Suppose the mass $m_d$ is increased continuously. As $m_d$ is dialed to $\infty$, the skyrmion gets continuously ``deformed" to a configuration in which $\eta^\prime$ winds around a singular ring~\cite{karasik}, which is  the FQH droplet of \cite{zohar}. This transformation requires  that
\be
\langle\chi\rangle\to 0, \ \langle\bar{q}q\rangle\to 0\label{dlfp}
\ee
and
\be
m_\omega\to0.\label{omega0}
\ee
In sHLS applied to dense matter, (\ref{dlfp}) corresponds to what is referred in \cite{PKLMR} to as ``dilaton limit" at which the axial current coupling constant $g_A$ in the neutron beta decay goes to 1.\footnote{As a side remark, I mention that this has a connection, though indirect, to the long-standing problem of   ``quenched $g_A$" in nuclear Gamow-Teller transitions~\cite{quenchedgA}. }As for (\ref{omega0}), in scale-symmetric HLS, the $\omega$ mass term is multiplied by $\chi^2$ by scale symmetry, so (\ref{dlfp}) makes the $\omega$ mass  go to zero.\footnote{The $\omega$-nuclear coupling itself is not required to go to zero as in the case of the vector manifestation for the $\rho$ mass.}

What transpires from relating the $N_f=1$ baryon to the $N_f\geq 2$ baryons as seen in the unified baryon current (\ref{gB}) could very well involve intricate topological inputs that have not yet been explored in nuclear physics. In order to incorporate such intricacy in CCP and to address consequent nuclear dynamics, one would require calculating the EoS of the baryonic matter ranging from normal nuclear matter density at which available experiments can be exploited  to high density for which neither reliable theory nor verified experimental data are available.  At present, such an EOS is missing. Given its absence, all one can do is inevitably to be adventurous and at best speculative. In what follows this is what I will do.

\section{From CCP to Dense Matter} \label{CCP2densematter}
It is not clear how to  zero-in on  the $N_f=1$ baryon and expose {\it directly} its FQH  structure of that baryon.  For a recent attempt to resolve this issue, see \cite{dichotomy}. I will simply assume that it intervenes, albeit indirectly, in baryon structure in nuclear matter under some (extreme) conditions and see what transpires.

 An intriguing possibility is its role in the proton spin. In various quark models such as in the MIT bag model and the constituent quark model, the flavor-singlet  axial-vector coupling $g_A^0$ is directly related to twice the proton spin, i.e.,  1. What was measured in deep-inelastic experiments, however,  was instead nearly zero. This was referred -- most likely wrongly~\cite{leader} -- to as the ``proton-spin crisis.  In the Skyrme model with the pion only that I will denote Skyrme$_\pi$,  $g_A^0 = 0$. This is actually closer to what was measured experimentally.  But from the fundamental theory point of view this  Skyrme$_\pi$ result cannot be correct.  In fact it turns out to come out correctly -- with $g_A^0\approx 0.3$ -- when the Chern-Simons term coupled to $\eta^\prime$ is taken into account on the bag boundary~\cite{gA0}. The Chern-Simons term figures in the CCP as a gauge-non-invariant boundary term to cancel the  color-anomaly induced quantum mechanically inside the bag~\cite{colorleakage}. This means that the FQH structure could actually play a basic role but only a small one in the  proton structure. Likewise the $O(N^0)$ contribution in the mass difference $\Delta M$ must  also be small, although perhaps not  zero. It seems also highly reasonable to assume that at normal nuclear density, the FQH  structure must play, if any,  an insignificant role.  Thus it seems safe to assume that the EoS at low density is negligibly influenced by the ``high-lying" flavor-singlet component in the nucleon structure.   Up to date  there is no indication of the presence of the flavor-singlet component  mixed into the nucleon structure in precision nuclear structure calculations.

In short, nature indicates that at low density in nuclear processes, the jail escape scenario seems to go predominantly  via the infinite-hotel mechanism into the skyrmions, and negligibly, if any, via the anomaly in-flow to the FQH droplet.

So whether and how it can  be ``seen" in nuclear physics is an open issue.

As shown in \cite{MR-review} and mentioned below, as density increases to high up, both the vector degrees of freedom (hidden local fields $\rho$ and $\omega$) and hidden scale symmetric field $\chi$ are found to figure importantly in massive compact stars. The dilaton condensate $\langle\chi\rangle$ and the quark condensate $\langle\bar{q}q\rangle$ are found to diminish, vanishing at some high density. The vector manifestation (VM) makes the massless $\rho$ meson decouple from the baryons and as mentioned, the vanishing dilaton condensate drives $m_\omega\to 0$. Furthermore at high density, the mass of $\eta^\prime$ does tend to decrease, making the $N_f=1$ baryon mass decrease.  Thus it seems reasonable  to expect  that the FQH structure, perhaps in the form of pita as will be suggested below,  could very well  play a role  in the  EoS  for massive neutron stars.
\subsection{Trading-in topology for quarks/gluons}
\subsubsection{Strategy: Topology change}
Let's start with the (u, d)-quark baryonic matter. Adopting the CCP, let's first consider the $\infty$-hotel structure, namely the skyrmion matter.  As explained above, incorporating the topological structure associated with the $N_f=1$ baryon necessarily requires both the vector mesons and scalar dilaton. Even at near nuclear matter density, there is a strong indication that those ``heavy" degrees of freedom could play an important role in the skyrmion description of nuclei and nuclear matter.  For instance, the binding energies and the cluster structure of light nuclei are much better described with them than with the Skyrme$_\pi$  model that contains only the pion degree of freedom~\cite{sutcliffe}. It's certain that the description of nuclear structure will improve markedly with the sHLS degrees of freedom included over what has been already achieved, which is quite impressive as it is, with the Skyrme$_\pi$ model~\cite{multifacet,mantonetal}.

In going to higher density, say, for $n\gsim 3n_0$, the impact of the sHLS degrees of freedom could become much more dramatic: They do not just improve over the Skyrme$_\pi$ results,  but they are in fact indispensable, even without possible impact of the FQH structure. Given a generalized skyrmion Lagrangian in which  sHLS is duly implemented, the problem is how to incorporate in doing full sophisticated many-body calculations the subtleties involved, e.g., the topology and singularity associated with the anomalies as encountered in going from (\ref{gB}) to (\ref{pancakeB}).  Unfortunately even at low density, near $n\sim n_0$, a systematic quantum and quantitatively manageable  treatment of the skyrmion structure in sHLS  is a formidable task not yet feasible. This will certainly be more the case for going to higher density.

At present, the only technique available -- and sufficiently trustful --  that one can resort to  is,  in spirit,  along the line taken in arriving at (\ref{gB}) from (\ref{pancakeB}) for unifying $N_f\geq 2$ and $N_f= 1$~\cite{karasik}. In going from (\ref{gB}) to (\ref{pancakeB}) in \cite{karasik}, certain topological inputs, crucial for the requisite ``deformation,"  enter in the parameters of the effective Lagrangian ${\cal L}_{\eta^\prime\omega\chi}$. In \cite{PKLMR},  essentially the same strategy has been used in accessing compact-star matter: Put appropriate topological inputs provided by CCP in the parameters of the nuclear sHLS Lagrangian and calculate the EoS by means of a suitable RG treatment. However the topological ramification that ``un-distorts" (\ref{pancakeB}) to (\ref{gB}) for the unified baryon current has not been explicitly implemented. Hence  that  the EoS  without such implementation seemed to work well for compact stars~\cite{PKLMR,MR-review} may be suggesting that possible contribution of the $B^{(0)}$  via  the FQH  structure, negligible at low density, could enter, in a ``metamorphosed" form, at the compact-star density.
 
 The question we address is: In what way the flavor-singlet nature of the proton structure, metamorphosed,  could figure in dense matter?
\subsubsection{Skyrmions on crystal lattice}\label{skyrmioncrystal}\label{qpstuff}
It has been suggested that potentially non-trivial information embedded in the topological structure of baryons could be extracted by putting skyrmions on crystal lattice. Skipping details which can be found in the review~\cite{park-vento}, let me list the key points involved.

When skyrmions constructed  with sHLS Lagrangian are put on FCC crystal lattice to simulate baryonic matter, the following results are obtained:
\begin{enumerate}
\item  {\bf Topology change}:
 A skyrmion (with $B=1$) in baryonic matter fractionalizes to two half-skyrmions (with $B=1/2$) at a density denoted  ``half-skyrmion density" $n_{1/2}$. In the half-skyrmion state, the quark condensate $\Sigma\equiv \langle\bar{q}q\rangle$ -- which is non-zero at low density  $n\leq n_{1/2}$ -- goes to zero when space averaged,  $\tilde{\Sigma}=0$, for n$>n_{1/2}$.  While zero space-averaged,  however, $\Sigma$ is locally non-zero, hence supports a chiral density wave. This implies that the pion decay constant $f_\pi$  is not equal to zero even though $\tilde{\Sigma}=0$, hence chiral symmetry is in the Nambu-Goldstone mode. Therefore the topology change involves no phase transition in the sense of the Landau-Ginzburg-Wilson paradigm. It may be likened to what's known as ``pseudo-gap phase" in condensed matter physics.\footnote{As mentioned later in a different context, this phase may be related to Georgi's vector symmetry.}  For convenience, I will (mis)use the terminology ``half-skyrmion phase." Where the density  $n_{1/2}$ is located  depends on the parameters of the sHLS Lagrangian. There is no known theory to fix it. The compact-star phenomenology  tuns out to give  $n_{1/2}\approx 3n_0$~\cite{MR-review}. 
\item {\bf Equation of state (EoS)}:
One of the most important observations in the skyrmion crystal analysis, playing the key role in compact-star physics,  is the cusp structure at $n_{1/2}$ in the symmetry energy $E_{sym} (n)$ in the EoS of the star matter
\be
E(n,\delta)= E(n,0)+\delta^2 E_{sym} (n) +O(\delta^4)\label{EoS}
\ee
with $\delta=(N-P)/(N+P)$ with $N(P)$ standing for the number of neutrons(protons). The symmetry energy $E_{sym}$ can be calculated in the crystal simulation by rotational quantization of the skyrmion matter as a whole, which comes at $O(1/N_c)$ in $1/N_c$ expansion in the energy density (\ref{EoS}), thus N$^2$LO in $N_c$ giving a cusp structure. It takes the form $E_{sym}\approx 1/(8\lambda_I)$ where $\lambda_I$ is the isospin moment of inertia~\cite{sym-LPR}. The key point in this formula is that the cusp at the skyrmion-half-skyrmion transition at $n_{1/2}$ is formed by  the ``heavy" degrees of freedom, principally the $\rho$ meson (in the Skyrme$_\pi$ model, the Skyrme quartic term), dominating over the pion contribution. As I will mention below, this cusp structure is ``softened" by the same HLS fields as the cusp in the $\eta^\prime$ potential in the chiral Lagrangian for $\eta^\prime$ is eliminated by the HLS fields in the FQH pancake baryon~\cite{karasik2}.
{\it It is a topological effect, and hence should be robust~\cite{robust}}. 

The  strategy in the approach is to implement this cusp structure  in the nuclear effective field theory incorporating sHLS, that I will call  ``generalized nuclear EFT" denoted $Gn$EFT. It enters in the change of the $\rho$-nuclear coupling $g_V$ which starts droping in strength as density increases beyond $n_{1/2}$. This impacts the nuclear tensor force $V_T$ in such a way that $|V_T|^2$ develops a cusp at $n_{1/2}$.  This cusp  is reflected directly in the symmetry energy, since $E_{sym}\propto V_T^2$.  This is one of the key points of exploiting the topology change for formulating $Gn$EFT. This property is encoded in the vector coupling $g_V$ contributing to $V_T$ as a function of density.  This behavior of  $g_V$ as a function density is intricately related to how the ``vector manifestation (VM)"  with $m_\rho\to 0$ is approached at high density. This result is also  intimately tied to the mass formula $m_\rho^2 \propto g_V^2$, which holds to all loop orders in HLS theory~\cite{HY:PR}.
\item {\bf ``Quasiparticles"}:
Another equally striking observation resulting from the cusp structure in $E_{sym}$ is the soft-to-hard change in the EoS at $n_{1/2}$. This is reflected in the symmetry energy that decreases in going toward $n_{1/2}$ from below and then increasing after $n_{1/2}$. The increase of $E_{sym}$ for $n > n_{1/2}$ is linear in density, which suggests nearly non-interacting ``quasiparticles" in the half-skyrmion phase. This can be interpreted as the hard-core repulsion between ``quasiparticles" at high density as observed in the constituent quark model supported by lattice simulation~\cite{cqm-lattice-repulsion}. This feature can be seen in the half-skyrmion phase {\it with or without}  the heavy-meson degrees of freedom in sHLS: It has been shown both in the skyrme model with the pions only (skyrmion$_\pi$)~\cite{atiyah-manton} and in the skyrmion model with sHLS fields (skyrmion$_{\rm sHLS}$~\cite{PKLMR}). {\it Note that this reflects that a robust topological effect is at work here.} The detail arguments are given in the two references given above. Let me just state the key observation involved using the skyrmion$_\pi$ model. This is reliable since what matters is the topology and it is the pion field that carries topology in the skyrmion models. 

What we need to calculate is the energy density of the skyrmion matter $\epsilon (n)$. In the skyrmion$_\pi$ model, that means solving the equations of motion for $\phi_0$ and $\phi^j_\pi$ of the chiral field 
\be
U(\vec{x})=\phi_0 (x,y,z) +i\phi_\pi^j (x,y,z)\tau^j
\ee   
in a skyrmion-crystal simulation. The field configurations $\phi_{0,\pi}$ correspond  effectively to the the mean fields in $Gn$EFT which transcribed into Fermi-liquid theory~\cite{PKLMR},   can be taken equivalent  to the Fermi-liquid fixed point quantity in the scale-HLS Lagrangian.   
\begin{figure}[h]
\begin{center}
\includegraphics[width=4.2cm]{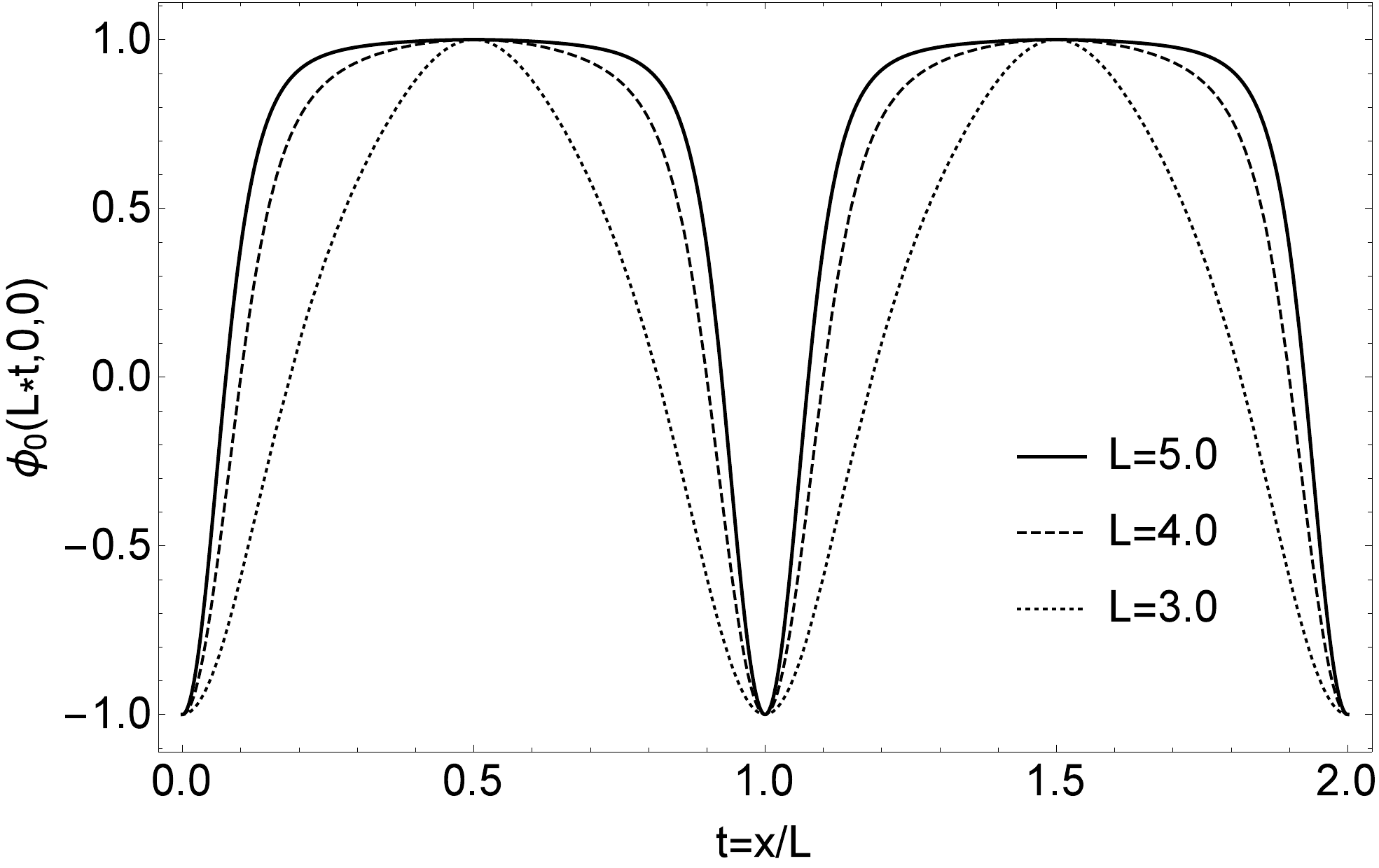}\includegraphics[width=4.2cm]{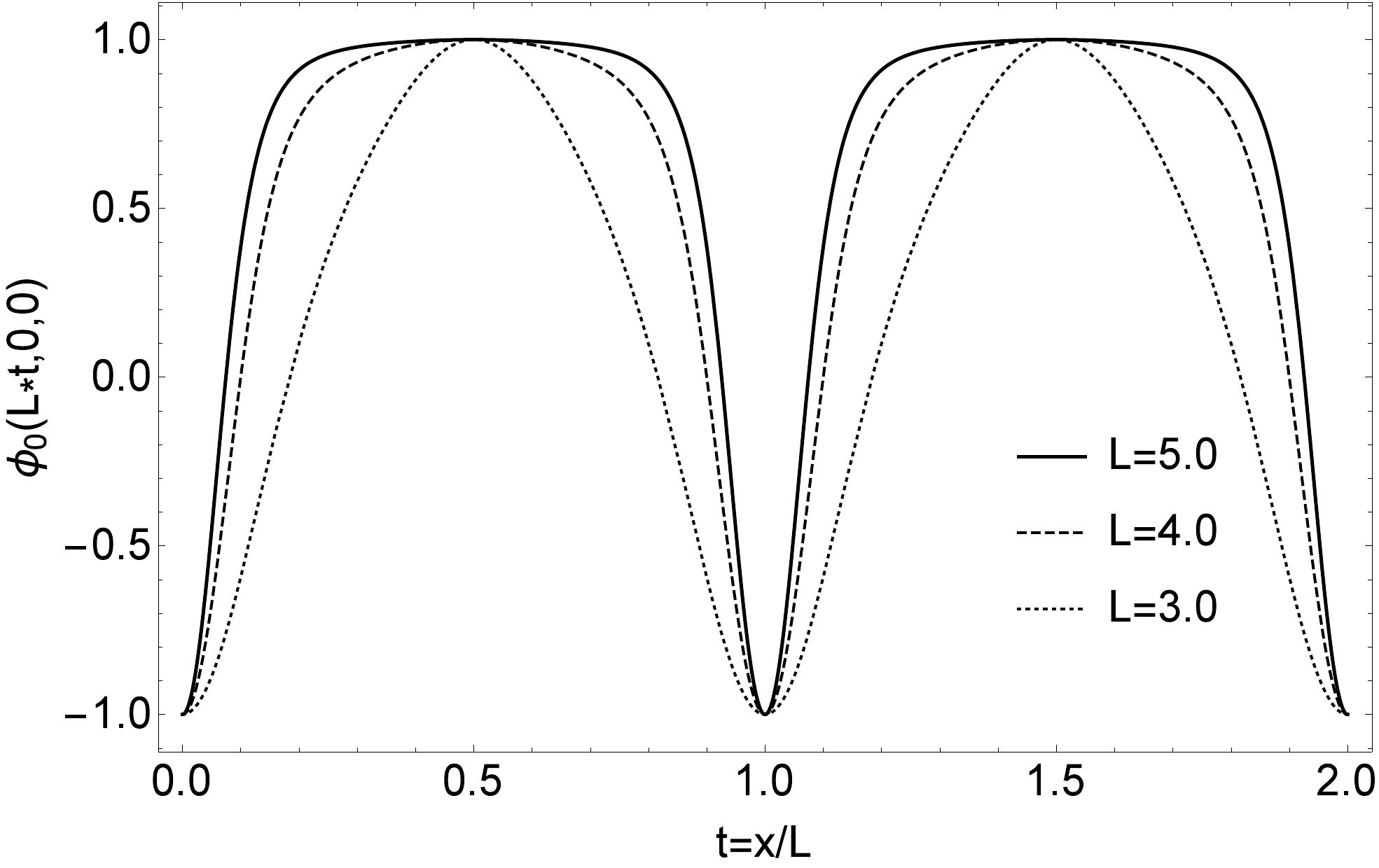}
\includegraphics[width=4.2cm]{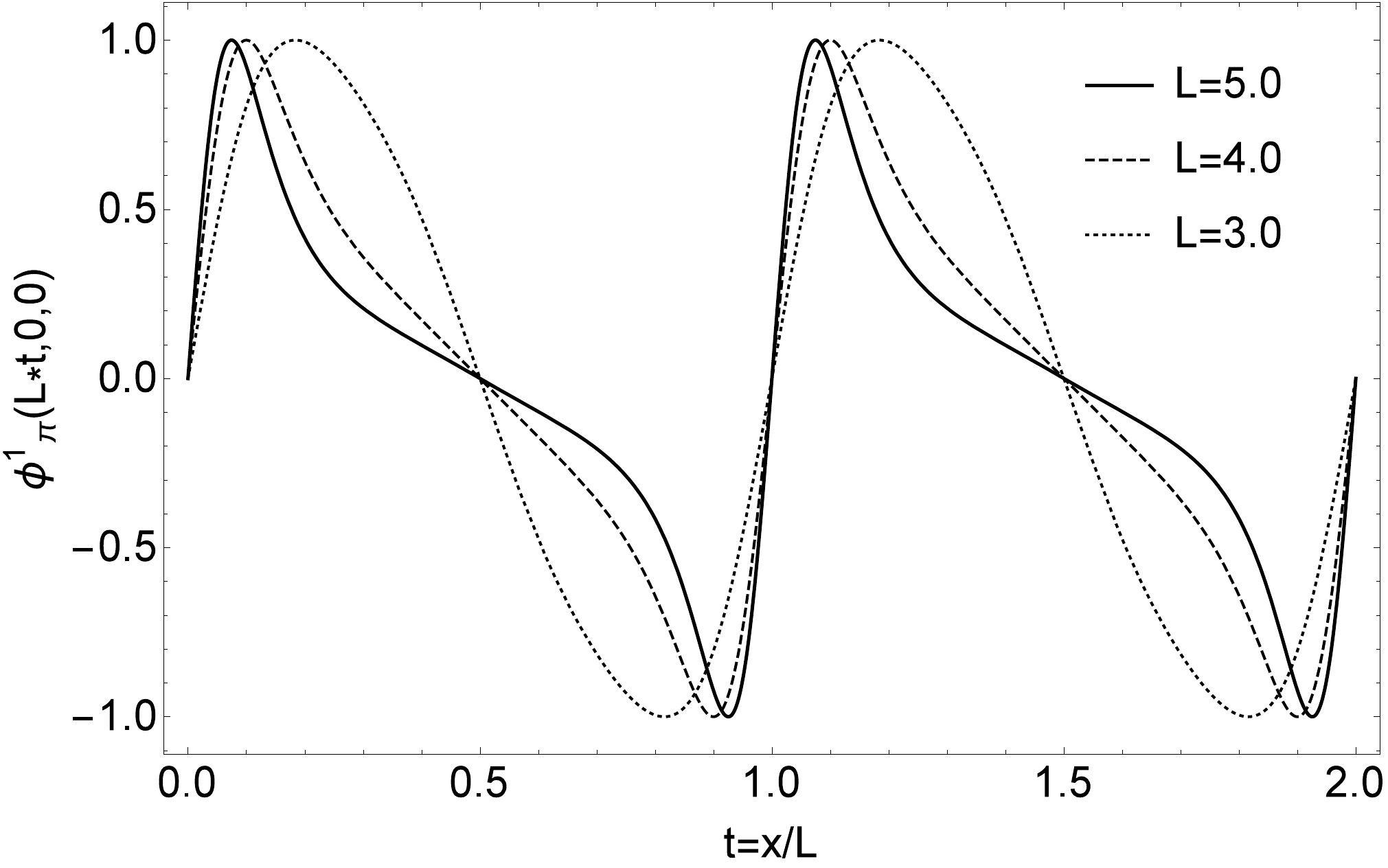}\includegraphics[width=4.2cm]{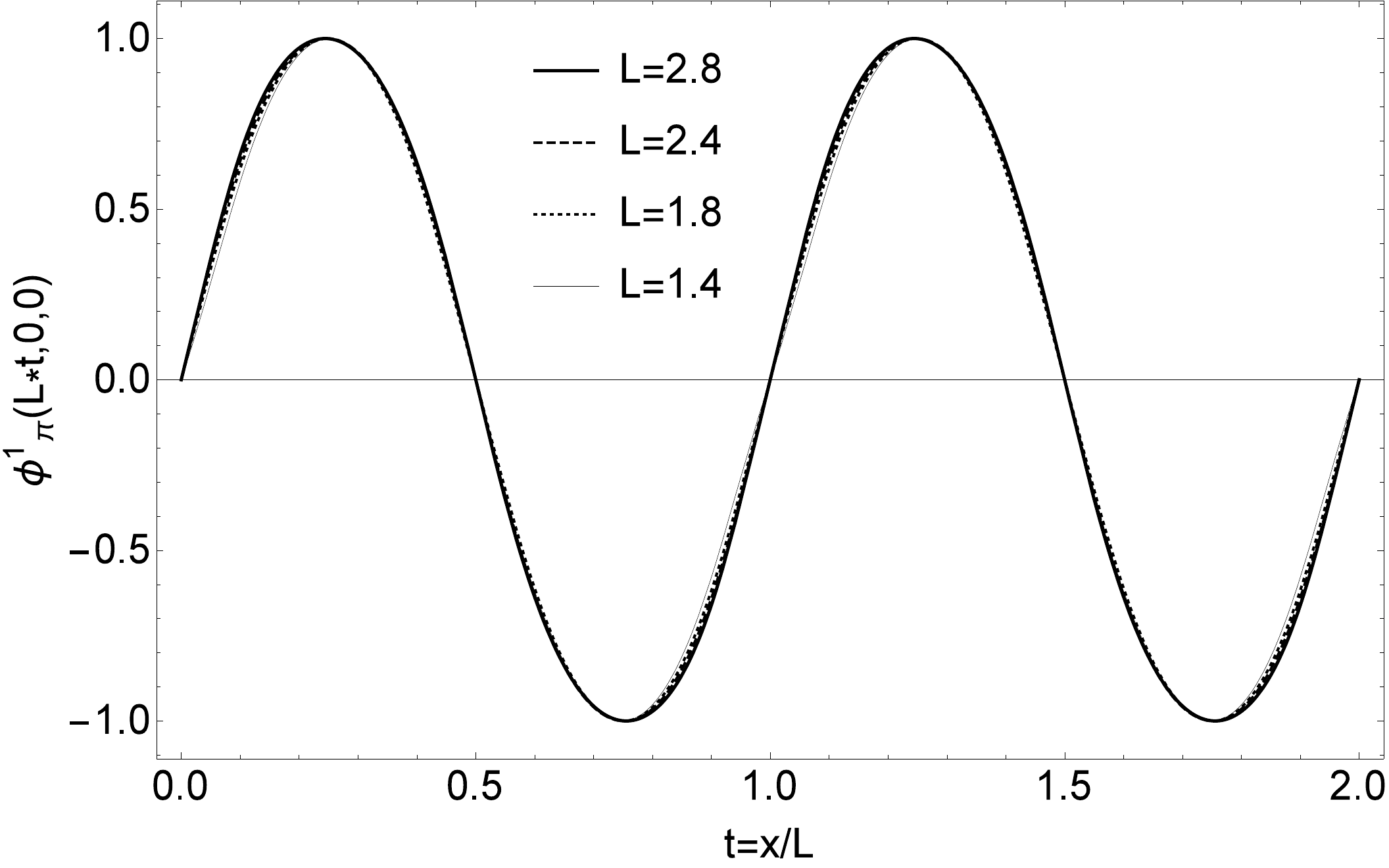}
\caption{  The field configurations $\phi_0$ and $\phi^1_\pi$  as a function of $t = x/L$ along the y = z = 0 line. The left panels correspond to  $n < n_{1/2}$ and the right panels to $n\gsim n_{1/2}$. The half-skyrmion phase sets in when $L=L_{1/2} \lsim 2.9\,{\rm fm}$. What is to be noticed is the density independence  of the configurations $\phi_0$ and $\phi^1_\pi$ which engenders scale invariance in the half-skyrmion sector
 }\label{scale_inv}
 \end{center}
\end{figure}

The  crystal simulations for the field configurations have been described in detail in two papers~\cite{atiyah-manton,PKLMR}. The results reproduced in Fig.~\ref{scale_inv}~\cite{PKLMR}, totally unrecognized in the field,  are striking.  Remarkably those configurations,  varying strongly due to  nuclear correlations with increasing densities in the skyrmion phase at $n < n_{1/2}$ become {\it density-independent in the half-skyrmion phase at $n\gsim n_{1/2}$}. {\it This means that those configurations representing non-interacting quasiparticles behave scale-invariantly.} This was reflected in the linear density dependence in the cusp for the symmetry energy in the half-skyrmion phase.

Particularly interesting is the density-independent configuration $\phi_0$. Since this quantity is more or less equal to to the pion decay constant $f_\pi$ -- and the dilaton condensate  $f_\chi$ gets locked to   $f_\pi$ going toward the IR fixed point in Crewther's  ``genuine dilaton" scenario~\cite{crewther}, this behavior of $\phi_0$ has impacts on two important quantities at high density,  first the sound velocity $v_s$ of compact-star matter discussed above
and second the possible Kohn-Sham energy density functional approach to topology discussed below.

\item{\bf Parity doubling}:
A feature in the skyrmion crystal simulation that plays also a very striking role in the strategy is the emergence of parity-doubling symmetry  in the nucleon spectrum for $n>n_{1/2}$~\cite{robust}. The skyrmion mass $m_{skyrme}$ which can be identified as the effective nucleon in-medium mass $m_N^\ast$ behaves as
\be
m_N^\ast\approx m_{skyrme} =\hat{\Sigma}+m_0\to m_0 \ {\rm as} \ n\to n_{1/2}.
\ee
Although $\hat{\Sigma}$ is not an order parameter for the NG mode of chiral symmetry  since the pion decay constant remains non-zero there, $m_0$ is a chiral scalar, a situation resembling the ``pseudo-gap" phase in superconductivity. It gives a parity doubling in the nucleon spectra in medium. Numerically $m_0$ is $\sim (0.6 -0.9)$ times the zero-density mass, so $m_0$ is substantial. Whether this is related to topology is not at all clear.  It seems intimately connected to the origin of the nucleon mass~\cite{origin}. 

The impact of this parity doubling is crucial in the EoS. With the dilaton field $\chi$ implemented (via sHLS) in the skyrmion crystal calculation, it is found as density is dialed to $n> n_{1/2}$ that 
\be
f_\pi^\ast\approx f_\chi^\ast\approx \langle\chi\rangle^\ast
\ee and 
\be
m_N^\ast\propto \langle\chi\rangle^\ast \to m_0
\ee
where the $\ast$ stands for density dependence.  At some high density referred to as ``dilaton limit fixed point (DLFP)" at, say, $\sim 25n_0$, $\langle\chi\rangle$ is expected to approach zero.   Below, this will be connected to the density regime where the flavor-singlet baryon degree of freedom becomes most likely relevant to the EoS as $\langle\chi\rangle$ approaches zero.
\item{\bf Quark-hadron duality}: 
It is suggested that what takes place involving the cusp structure is a trade-in of topology for putative quark-hadron continuity in the EoS as hadrons change over to quarks at increasing density~\cite{baymetal} and that the half-skyrmion matter is dual,  via the CCP, to that of quarks/gluons in nonperturbative and highly correlated interactions~\cite{MR-review}.  At present, it should be admitted, there is no rigorous argument for this.  Nonetheless many of the features so far deduced, albeit semi-quantitatively,  resemble what has been obtained in some interesting models where quarks figure explicitly, e.g., \cite{quarkyonic}. In both cases there is no phase transition involved in going from hadrons to quarks. 
\end{enumerate}

\subsection{Nuclear effective field theory with sHLS: $Gn$EFT}
Instead of attempting to formulate directly the CCP to unify  the descriptions of  the  $N_f=1$ baryon (FQH droplet) and the $N_f\geq$ baryons (IH skyrmions) in the EoS, which is beyond our capability, I will take the strategy similar to what has been taken by Karasik for the baryon current in terms of the unified baryon current that incorporates both the HLS mesons $(\rho,\omega)$ and scalar dilaton ($\chi$). As in the EFT Lagrangian currently popular in nuclear theory, nucleons are explicitly included. To access the EoS for dense matter, the strategy then is to implement the degrees of freedom of sHLS to a nuclear effective Lagrangian denoted ``$Gn$EFT" by incorporating both the topological features and non-topological features listed above that are extracted from the skyrmion crystal analysis. {\it Those features are let to control  the parameters of the $Gn$EFT Lagrangian}. 

Now the question is how to approach, with the given $Gn$EFT, many-baryon dynamics  exploiting the robust topological inputs in accessing dense matter. The familiar (chiral) power expansion employed in standard chiral perturbative approach (to be denoted S$\chi$EFT), successful at low density, will become powerless at high density. This is more so with the heavy degrees of freedom present in the EFT Lagrangian. The approach adopted in \cite{PKLMR} is along the line of the Wilsonian RG approach to correlated fermions in the presence of a Fermi sea. The $Gn$EFT Lagrangian with the topological inputs plus the ``intrinsic density dependence (IND)" inherited from QCD at the chiral scale, when treated in the mean field, corresponds as first suggested by Matsui for Walecka's relativistic mean-field model~\cite{matsui}, to Landau Fermi-liquid  theory at the Fermi-liquid fixed point~\cite{polchinski,shankar}.\footnote{The fixed-point is in the limit $\bar{N}\equiv  k_F/(\Lambda-k_F)\to\infty$ where  $\Lambda$ is the cutoff on top of the Fermi surface.} Corrections to $O(1/\bar{N})$ to go beyond the fixed point approximation can be made in  the $V_{lowk}$RG approach. Both with and without $O(1/\bar{N})$ corrections  have been studied in \cite{PKLMR,MR-review}. It should be stressed that the density-dependent topological inputs plus the IND in $Gn$EFT Lagrangian render this energy density functional potentially more powerful than the standard RMFT available in the literature. 
%
\subsubsection{Normal nuclear matter}
Up to the topology change density $n\sim n_0$, $Gn$EFT is essentially equivalent to S$\chi$EFT -- say, to N$^4$LO with the same quality of fits to data~\cite{MR-review}.  It is expected that there will be little difference in predictive power at least up to $\sim 2n_0$. There is however one striking prediction in $Gn$EFT that has not yet been revealed in S$\chi$EFT -- as mentioned above and discussed in \cite{quenchedgA} -- on the phenomenon of ``quenched $g_A$" in allowed Gamow-Teller transition which has defied nuclear theorists for over four decades. This puzzle  seems to be resolved when the hidden symmetries are properly taken into account. It has been suggested that this may have a crucial impact on the problem concerned here~\cite{WS-invited}. 
\subsubsection{Massive compact stars}
Although continuous with no genuine phase transition, the matter at $\gsim n_{1/2}$ undergoes a drastic change due to the skyrmion-half-skyrmion transition  as already indicated. 
One of the most significant predictions  is the cusp in the symmetry energy $E_{sym}$ at $n=n_{1/2}$ that causes the EoS, soft at lower density, to become stiff at high density,  $\gsim n_{1/2}$, qualitatively simulating the soft-to-hard transition at $n\sim (2-4) n_0$ driven by  strong-coupled quark interactions in models simulating hadron-quark continuity~\cite{baymetal,quarkyonic}. This prediction made by the topology change that accounts for the massive compact stars of mass $\gsim 2M_\odot$ is extremely neat although it involves a highly intricate mechanism. 

When the putative hadron-quark continuity is approached by hybridizing -- with different, say, Lagrangians -- the standard nuclear EoS valid up to $\sim 2n_0$ to a ``quarkish" EoS for $n\gsim 2n_0$ to implement QCD degrees of freedom at high density, there tend unavoidably to be as many alternatives as there are theoretical efforts, some totally disconnected and some pretending to model  a trustful theory, inevitably unconnected to QCD proper. At present, the general tendency is that whatever new observables do appear in the literature that bring tensions between the model and experiments, can be fudged within the hybrid structure by a battery of new parameters to fit the data.  And most of the time they seem to work!   But the problem is that it is difficult to see in such treatments what it means to ``rule out" or ``confirm" any given EoS, given the ambiguities in the fitting process.

In a stark contrast, there are striking features associated with the topology-driven phenomena in the topology-change mechanism that are {\it not} shared by the hybrid baryon-quark-models with the merit that they can be unambiguously confirmed or ruled out. This is a unique power of the approach although it could be admittedly over-simplified or even downright wrong.

Most notable, it turns out,  is the role of the scalar dilaton  $\chi$ and the vector meson $\rho$ in the structure of massive compact stars with $M\gsim 2 M_\odot$.
%
\subsubsection{Pseudo-conformal sound velocity}

There are two issues recently studied that illustrate the uniqueness of the properties of the approach, whether right or wrong or can be improved. The first is the sound speed to be discussed here and the second is the ``stuff" in the core of the star which will be treated in Sect. \ref{core} below. As for other star global properties including recent gravity-wave observations, there is nothing so far glaringly at odds with the observations\footnote{There is a loud debate going on between astrophysicists -- and also among model builders -- on possible stars of mass $M\sim (2.3-2.67) M_\odot$ (see, e.g., \cite{>2.5star} among a humongous number of articles in the literature). If such massive stars turn out to be confirmed as neutron stars, then the present topology-based hadron-quark continuity might get in serious tension with the observation.}, so I won't go into them. (See \cite{MR-review}.) 

\begin{figure}[htbp]
\begin{center}
\includegraphics[width=0.5\textwidth]{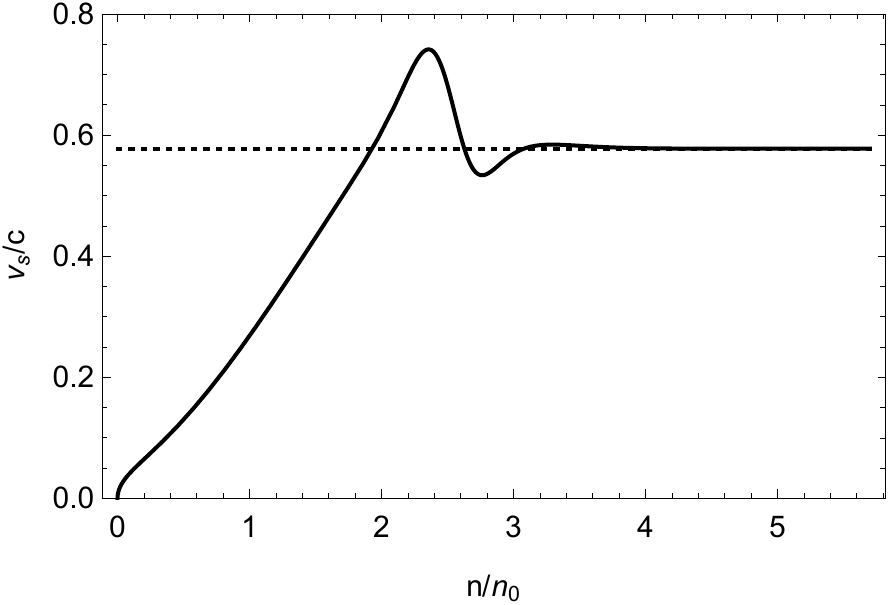}
\includegraphics[width=0.5\textwidth]{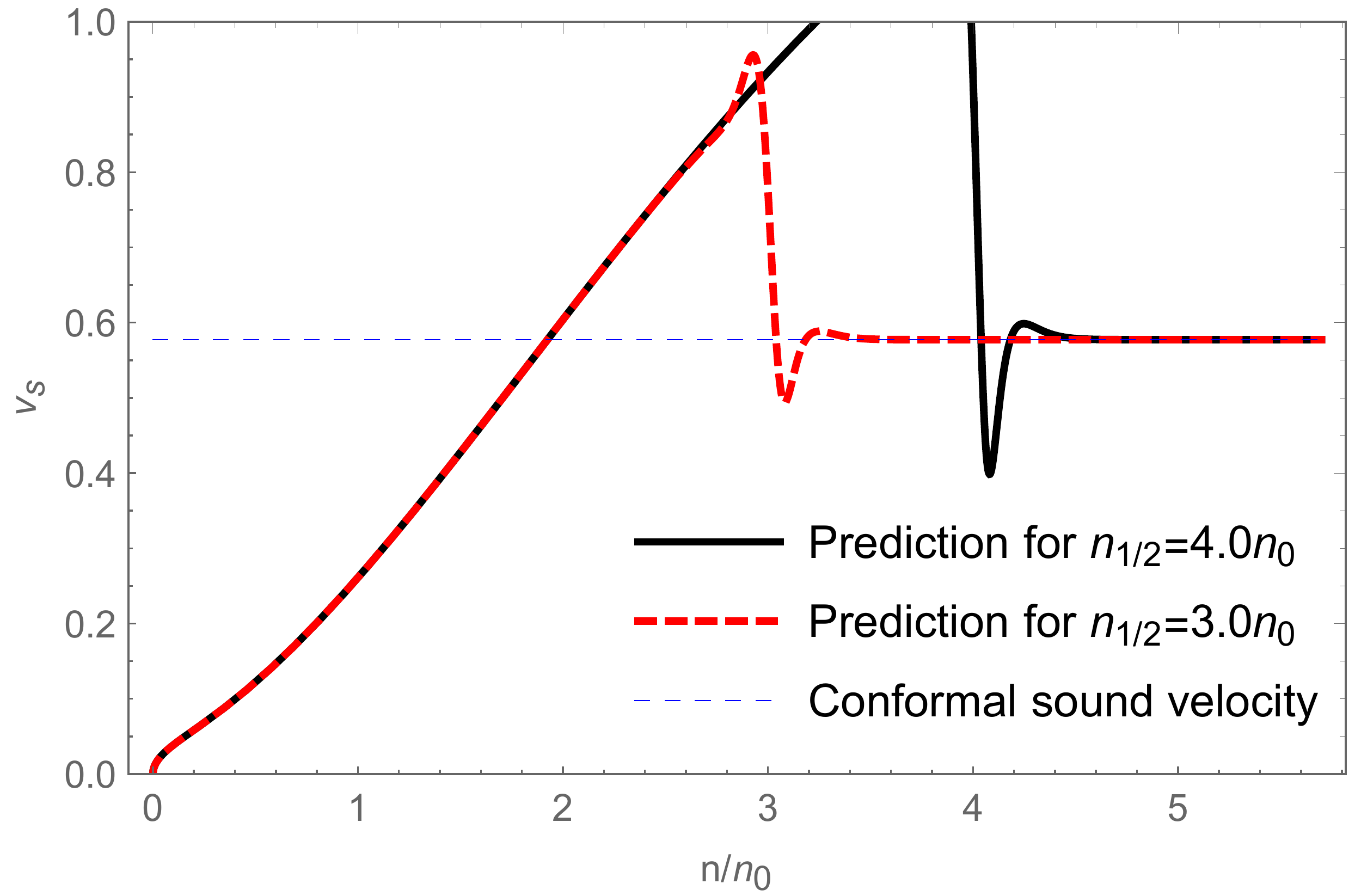}\end{center}
\caption{Density dependence of the sound velocity $v_s$  in neutron matter for $n_{1/2}/n_0= 2.5$ (upper panel) and $n_{1/2}/n_0= 3$ and 4 (lower panel). To be noted that for $n_{1/2}=4n_0$, the sound velocity violates causality at  $n\gsim 3n_0$ and becomes causal at $n\sim 4n_0$.  The sharpness of the peaks near $n_{1/2}$ could simply be an artifact of the way the topological changeover takes place, which is the most uncertain region of the theory.}
\label{vs}
\end{figure}

Surprisingly the structure of the sound speed $v_s$ for $n>n_{1/2}$ is drastically different for different VM fixed points $n_{VM}$s for the $\rho$ meson. If one were to take $n_{VM}\sim 6 n_0$, the density relevant to the core of neutron stars,  that  most of the standard nuclear models favor for chiral restoration (in the chiral limit), the sound velocity would increase monotonically from $v_s\sim 0.5$ at at $\gsim 2n_0$ to $v_s\sim 0.8$ (in units of $c=1$) at near $n_{VM}$. Details differ in standard nuclear EFT models, however, the models involving the {\it baryons only} as relevant degrees of freedom including density functional theories such as RMF theories tend to have this feature. In stark contrast, however, with $n_{VM}$ taken to be much higher than what's relevant to stable compact stars, $n_{VM}\gsim 25 n_0$, the $v_s$ converges to,  and stays at, the conformal velocity $v_s=\sqrt{1/3}$ from $\sim 3 n_0$ up to the interior of massive stars, $\sim (6-7)n_0$ and most likely beyond $\sim 10n_0$. This is related to the fact that in the former case, the trace of the energy momentum tensor (TEMT) $\langle\theta_\mu^\mu\rangle$ -- which is a function of $\langle\chi\rangle$ --  is density-dependent,  whereas in the latter case the TEMT -- which is not zero -- is independent of density. One finds for  ${n_{VM} \gsim 25n_0}$  
\be
\frac{\del}{\del n}\langle\theta_\mu^\mu\rangle|_{n>n_{1/2}}=0\ {\rm for}\  \ 3\lsim n/n_0\lsim 8.
\ee 
The result is that the sound velocity converges to $v_s^2=1/3$.  This is shown in Fig.~\ref{vs} (upper panel).

One cannot say what happens for $n\gsim 8 n_0$. Since $\la\theta_\mu^\mu\ra\neq 0$,  $v_s^2=1/3$ does not imply conformal invariance. It is, strictly speaking, {\it not} ``conformal velocity."  The present interpretation is that  it reflects a simulated scale symmetry ``emerging" from strong nuclear correlations at high density, not necessarily connected to a putative infrared fixed point in QCD proper. Therefore it is referred to as  ``pseudo-conformal (PC) velocity" associated with emergent pseudo-conformal symmetry~\cite{MR-review}.   What we have here is the result from $Gn$EFT but it is already indicated from the scale-invariant ``quasiparticle" structure of the half-skyrmion phase already seen in Section \ref{skyrmioncrystal}. What we see here is that the quasiparticle notion as is familiar in Landau Fermi-liquid structure of electrons in condensed matter and of nucleons in nuclear physics at low density -- at the Fermi-liquid fixed point -- therefore applies to high density, approximately of course,  where the fermions are neither pure baryons nor pure quarks, an aspect relevant to the constituents in the core of stars to which we will turn below.

How the pseudo-conformal velocity goes over to the  genuine conformal velocity expected at  asymptotic densities is not understood.
Furthermore why this highly surprising dependence of the sound velocity on the density at which local gauge symmetry is restored, $n_{VM}$,  remains un-understood. If it is not in error, it  must involve an interplay between the two hidden symmetries that begs to be deciphered. What seems obvious is that whatever is in action, it is quite analogous to the emergent scale symmetry captured in the $g_A$ at $\sim n_0$ via nuclear correlations as it becomes genuine when the dense matter moves toward the dilaton limit fixed point with $g_A^{DL}=1$~\cite{WS-invited}.
\subsubsection{``Much ado about nothing"...}
Needless to say,  a various approximations are made in the calculation made. For instance,  the density dependence resulting from the symmetry breaking effects (such as $\beta^\prime$,  the anomalous dimension of the gluon stress tensor) is ignored.  This was referred to as ``LOSS" approximation in \cite{MR-review}. This is essentially what's done in all nuclear physics,  in S$\chi$EFTas well as  density-functional-type approaches  although some of their effects may be subsumed in the parameters of the Lagrangians used. Furthermore in the approach I am adopting there could very well be higher-order $1/\bar{N}$ corrections to the Fermi-liquid fixed point approximation that could  bring additional density dependence. Thus the PCM prediction cannot be taken to be quantitatively accurate. It seems natural to expect fluctuations of various strength depending on details of approximation on the curve $v_s^2$ vs. density for $n\gsim n_{1/2}$. In particular the crossover from the ``hadronic" sector to the putative ``quark" sector must be complicated given that there is absolutely no reliable theory available in that region. It cannot be used to differentiate the quality of the descriptions. One can clearly see this in various recent publications addressing recent astrophysical  observables relying on various forms of sound velocity as inputs~\cite{v_s}. In what way this exploitation of the varieties of sound velocity probes the EoS of dense compact-star matter is not at all clear. This aspect can be illustrated in the PCM by varying the topology change density from $n_{1/2}=2n_0$ to $n_{1/2}=4n_0$ as plotted in Fig.\ref{vs} (lower panel).  While there is nothing special with the ``bump" in the transition region, apart from the size of the bump,  for $n_{1/2}<4n_0$, the $v_s$ for $n\gsim 4n_0$ strongly violates causality before and across the crossover. Despite this violent behavior,  practically all observable global star properties do not seem to differ  drastically within the range of $n_{1/2}$. There are some small  (and expected) differences for the maximum star mass ranging from $\sim 2.0 M_\odot$ to $\sim 2.3 M_\odot$ -- and similarly in the central densities -- but there is nothing drastic in other quantities such as the star radius, tidal deformability  etc.  comparable to  the ``bump" structure in $v_s$ in the vicinity of the crossover region. This means that possibly the drastic difference in complicated structure in the sound speed as discussed in the literature with various model EoSs~\cite{v_s} cannot be used to gauge the goodness or badness of EoS.   ``{\it Much ado about nothing}"? 

This criticism applies~\cite{sound/core,fraga} to certain class of hadron-quark hybrid structure currently being constructed microscopically -- including shell-like structure.

\subsubsection{The ``stuff" in the core of massive compact stars}\label{core}
Fortunately some recent developments indicate that further astrophysical observations and refined theoretical  {\it ab initio} calculations  could address the issues relevant to the predictions made in the PCM in $Gn$EFT formalism. Let me mention one recent work zeroing-in on this matter.

\begin{figure}[htp]
\begin{center}
\vskip 0.3cm
\includegraphics[width=0.4\textwidth]{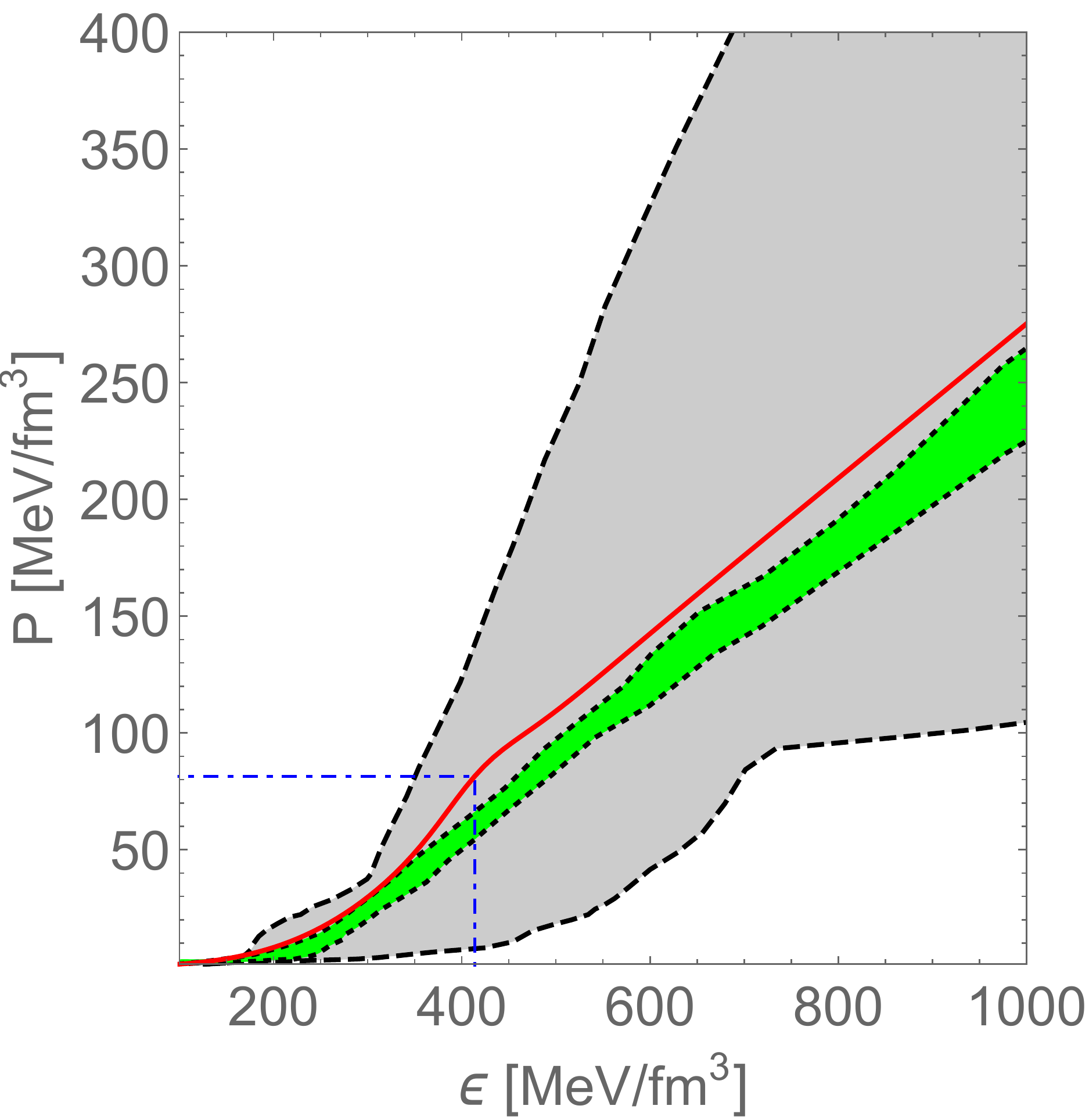}
\end{center}
\caption{Comparison of $(P/\epsilon)$  given by the band generated with the
sound velocity interpolation method used in~\cite{evidence}.  The green band is from the conformality~\cite{evidence} and the gray band from the causality.  The red line is the prediction of the pseudo-conformal model~\cite{cores}. The dash-dotted line indicates the location of the topology change.}
\label{P/E}
\end{figure}

By combining astrophysical observations spurred by recent gravity-wave data and theoretical calculations anchored on {\it ``ab initio"} approach, Annala et al. in the recent Nature Physics article~\cite{evidence} argued that the matter in the core of maximally stable neutron stars ``exhibits" characteristics of the state populated by ``deconfined" quarks.  As stressed by the authors, it is quite likely premature to come to any firm conclusion based on the rather involved analysis. Their analysis is captured in  the  sound speed approaching the conformal value $v_{sconf}^2=1/3$ as indicated by the green band in Fig.~\ref{P/E} and the polytropic index defined by $\gamma=d({\rm ln P})/d({\rm ln}\epsilon)$ going toward $\gamma=1$ at energy densities $\epsilon\approx 400-700$ MeV fm$^{-3}$, corresponding to typical energy densities for the deconfinement at high temperature in relativistic heavy ion collisions. The suggestion was that  the presence of ``deconfined quarks" in the core of massive stars should be the ``standard scenario and not an exotic alternative."  In addition there are arguments that other {\it banal} alternatives with large amplitudes (``bumps') in the range of densities relevant to massive stars~\cite{v_s} could be dynamically unstable and should be discarded~\cite{fraga}.

Let me now compare the PC model predictions with the results of \cite{evidence}.    

\begin{figure}[htbp]
\begin{center}
\includegraphics[width=0.4\textwidth]{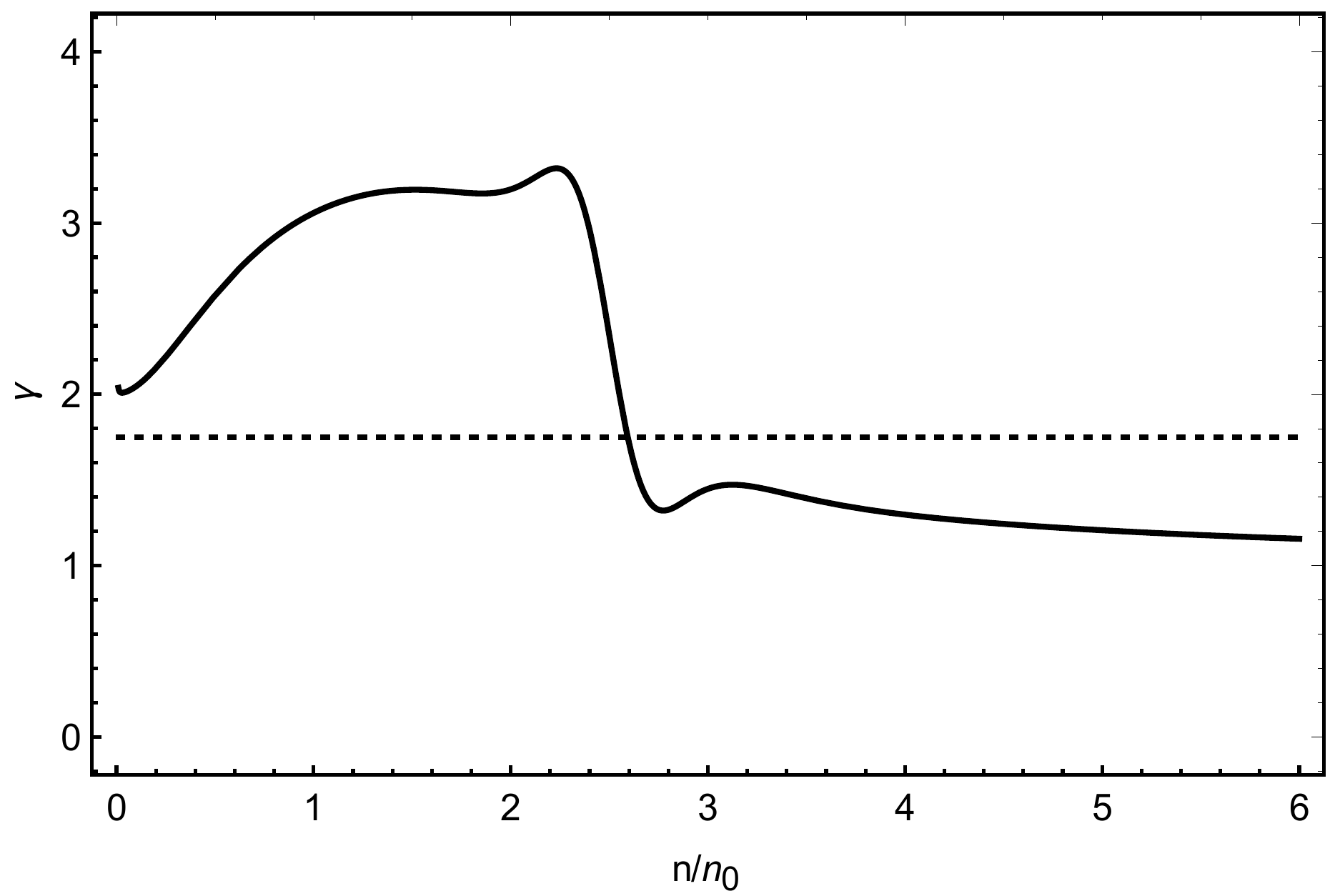}
\end{center}
\caption{Density dependence of the polytropic index $\gamma$ (in neutron matter.}
\label{PI}
\end{figure}

The PCM prediction for the sound speed is given in Fig.~\ref{vs}. The sound velocity approaches $v^2_{psv}=1/3$ for $n\gsim n_{1/2}$. 
The polytropic index $\gamma$ is  plotted in Fig.~\ref{PI}. It changes from $\gamma_{\rm nucl}\sim 30$ to $\gamma_{\rm pQCD}\sim 1$ in the changeover region as it does in \cite{evidence}. It should be noted that the predicted $P/\epsilon$  is close, and parallel, to the conformality band of \cite{evidence} as seen in Fig.~\ref{P/E}.
Thus,  at first look they seem to resemble closely, if not identical to,  the results of \cite{evidence}.

But there is a basic difference between the two. First of all  the PCM velocity is not  ``conformal" since the trace of energy momentum tensor (TEMT) cannot be zero. Although $v_s^2$ is close to 1/3, the $P/\epsilon$ must  deviate from the conformality by  density-independent constant terms in the TEMT predicted by the PCM. This feature can also be seen in the polytropic index. 
Conformality would imply that $\gamma$ is strictly 1 but one notes there is a small deviation from 1 at the density where $v^2_{s}$ approaches 1/3. Thus we see that the stuff in the core of massive neutron stars {\it cannot} be ``deconfined" quark-matter. Furthermore it cannot be purely hadronic  either. 
Since there is no phase transition involved, it must be  a new state of matter that is neither baryonic nor quarkonic. 

 So what is the ``stuff" that masquerades like a giant blob of deconfined quarks? 

My suggestion is that the resolution of the dichotomy problem mentioned above could provide an understanding of what's actually going on here.

\subsection{Density functional approach}\label{Kohn-Sham}
I now turn to what is included in $Gn$EFT that gives the results mentioned above and what seems to be needed in addition for resolving the dichotomy problem. This part is an ongoing work, so I can be at best speculative.

The topological structure that was mapped to $Gn$EFT in \cite{MR-review} was the half-skyrmion phase simulated on crystal lattice. It seems to have worked so far without getting at odds with nature. 

The question I want to address now is: How can one incorporate the mechanism that brings ${\calB}_{N_f=1}$ and ${\calB}_{N_f\geq2}$ into ${\calB}_{unif}$  in nuclear dynamics at  densities ranging from low to high? Stated differently, how to ``unify"  the $\infty$-hotel and the FQH structures  in EoS?  If this is feasible,  how important is the latter for compact-star physics? 
\subsubsection{Mapping to topology to $Gn$EFT}
The way to answer the  question raised  must involve translating the topological strategy in bringing ${\calB}_{N_f=1}$ and ${\calB}_{N_f\geq2}$ into ${\calB}_{unif}$ into the strategy of $Gn$EFT. A recent new development relevant, perhaps more conceptually to this matter, is the work treating the fractional quantum Hall  (FQH) phenomenon in the Kohn functional density approach \`a la  Kohn-Sham~\cite{jain}.  It corresponds to ``mapping" between the Chern-Simons field theory, powerfully exploited in FQH effects~\cite{tong}, and Kohn-Sham density functional theory.  
In this mapping~ \cite{jain}, the key ingredient is the weakly interacting composite fermions (CF) formed as bound states of electrons and quantum vortices. They are treated in Kohn-Sham density functional theory to arrive at the FQH states that capture  the strongly-correlated electron interactions encoded  in the topological field theory. Now in quantum Hall physics, the Kohn-Sham approach can be considered as a microscopic approach to the strong electron correlations involved in terms of the wave functions, e.g., Laughlin wave function, as contrasted to  Chern-Simons topological field theory which is a coarse-grained macroscopic approach.

Given that the same  Kohn-Sham theory in nuclear physics more or less underlies practically {\it all} nuclear EFTs employed with success in nuclear physics at low density, as for instance  energy density functional approaches to nuclear structure, and our $Gn$EFT approach, as described~\cite{MR-review},   belongs to this class of density functional theories extended to higher densities with the hadron-quark duality implemented, it seems therefore tempting to approach the {\it dichotomy problem}, or more generally hadron-quark continuity (or duality),  in a way similar to what was done in \cite{jain} for FQHE.  Thus what is to be done is to implement the dichotomy structure described above into the density functional $Gn$EFT Lagrangian in accessing high density regime.  Unfortunately  the microscopic physics is totally different between the two, so there is no close parallel to rely on for intuition. Also what's addressed in FQH effect is the fine-grained structure of excitations involved whereas in compact-star physics, it is the EoS and global coarse-grained properties associated with it.  
\subsubsection{A speculation}
There are some elements that seem to share similar features between the FQHE in condensed matter and the dense baryonic matter.  For instance  $Gn$EFT anchored on the topology change could be capturing the weak CF structure of \cite{jain} in FQHE,  crucially relying on the nearly non-interacting ``quasiparticle" behavior stressed above in the chiral field configuration $U$ in the half-skyrmion phase discussed in Section \ref{skyrmioncrystal}  (see Fig.~\ref{scale_inv}).  One way to understand this feature is as follows:  Due to $U(1)$ gauge symmetry in the hedgehog configuration of skyrmions, the half-skyrmion carries a magnetic monopole associated with the  hidden $U(1)$ symmetry in in the chiral field $U=e^{i\pi/f_\pi}$~\cite{cho}.  The energy of the ``bare" monopole in the half-skyrmion diverges when separated, but the divergence is tamed by interactions, as obvious in the skyrmion,  as a bound state of two half-skyrmions. In a way analogous to what happens in the Kohn-Sham theory of FQHE~\cite{jain}, there  could  intervene the gauge interactions between the half-skyrmions in sHLS-- as composite fermions -- possibly induced by the Berry phases due to the magnetic vorticies. Thus it seems plausible that the topological structure of the FQH is buried  in the half-skyrmion structure with the effect of the $B^{(0)}$  structure captured at a higher density. 

Another point of interest is that in making a link to the density functional approach for dense matter, it is an interesting question whether the $N_f=1$ baryon is a FQH pancake~\cite{zohar} or a  pita which can be thought of a  pair of pancakes with 1/2-charged edge modes~\cite{karasik}. For the pita configuration,  $\langle\bar{q}q\rangle=0$ but $f_{\eta^\prime} \neq 0$. Now recall that the half-skyrmion phase is characterized by $\langle\bar{q}q\rangle=0$ but $f_\pi\neq 0$. There is thus an  analogy between the case of $N_f=1$ and $N_f=2$, both involving 1/2-charged objects.\footnote{This could be interpreted in terms of Georgi's ``vector limit" (VL)~\cite{georgi} instead of the vector manifestation limit~\cite{HY:PR}. If the VL is involved, then one can expect that $f_S/f_{\eta^\prime}=1$ and $f_{\eta^\prime}\neq 0$. In matter-free space, the Georgi vector limit was ruled out by Ward identities~\cite{HY:PR} but in the presence of dense medium, this no-go theorem may not hold.}

As noted, at low density, the quarks in the bag would tend more likely to fall into the infinite hotel, hence giving rise to skyrmions in (3+1)d.  This may be driven by the parameters of the Lagrangian that unifies the $N_f\geq 2$ and $N_f=1$ baryons in $\calB_{unif}$ to have the $B^{(0)}$ structure unstable or {\it suppressed} at low density.  However as density increases,   the parameter change in the sHLS Lagrangian that distorts the baryon current ${\calB}_{unif}$ to ${\calB}_{N_f=1}$ could transform the EoS state toward a density-functional FQHE. 

One possible scenario for this is indicated in the recent skyrmion crystal analyses of dense matter where an inhomogeneous structure is found to be energetically favored over the homogeneous one at high density.  It has been found that dense matter consists of a layer of sheets of ``lasagne" configuration with each sheet supporting half-skyrmions~\cite{PPV}.\footnote{There is also a numerical analysis in the Skyrme$_\pi$ model which finds a stack of tubes supporting $1/q$-charged skyrmions where $q$ is an odd integer~\cite{canfora}.} The constituents of this layer structure are fractionalized quasi-stuff of 1/2 baryon-charged objects, appearing in baryon-quark continuity at a density $\sim n_{1/2}$. This is of course  drastically different from the pasta structure discussed for the dilute outer layer of compact stars. In the Skyrme$_\pi$ used here,  the quartic (Skyrme) term effectively encodes massive degrees of freedom,  including hidden local fields, monopoles  hidden in half-skyrmions etc.  It seems feasible to formulate this ``sheet dynamics"  by a stack of FQH-type  pitas, with tunneling half-skyrmions between the stacks{\footnote{Somewhat like arriving at the Chern-Simons structure of FQHE in (2+1)D with a stack of quantum wires~\cite{quantumwires}}. 
%
%

In discussing the quasiparticle properties of the half-skyrmion phase, the half-skyrmions are taken to be bound or confined and the pair behaves as a quasiparticle associated with Landau Fermi-liquid structure.  Whether or not they possess the characteristics of Landau Fermi liquids as in the electron systems cannot be addressed in this description. The notion of Landau Fermi-liquids in the renormalization-group approach to strong fermionic correlations~\cite{shankar} valid at near normal matter density $\sim n_0$ is assumed to extrapolate to high density with the parameters of the Lagrangian modified by topology change defined on the Fermi surface, hence the notion of quasiparticles,,  goes over from low to high density. This notion must however break down when  the dilaton-limit fixed point  is approached as  it does for Fermi liquids approaching the unitarity limit~\cite{nonfermi}.  As already stressed, however the dense compact-star matter we are concerned with is a distance away from the GD (genuine dilaton) fixed point, thus the Fermi-liquid structure as reflected in the PC symmetry could be valid~\cite{nonfermi}. 

In standard pictures such as the constituent quark (a.k.s. quasiquark)  model, the relevant fermion degrees of freedom are 1/3-baryon-charged. In the application to compact stars in \cite{MR-review}, the relevant fermionic charge was 1/2 -- but neither 1/3 nor deconfined.  In microscopic approaches to hadron-quark continuity where S$\chi$EFT or similar standard nuclear model is hybridized to quark models, it is the constituent quarks that figure in the higher density regime. In the topology-charge approach without the dichotomy problem as in \cite{MR-review}, it is the  bound pairs of 1/2-baryon-charged objects that figure.  With the dichotomy resolved, it could be that the stack of sheets with the Chern-Simons fields (metamorphosed from HLS) could be populated by fractionally charged ``deconfined" objects analogous to quantum magnets with ``deconfined spinons" on domain wall~\cite{deconfinement-wall}.  On the FQH pita, the bound 2 half-skyrmions could be liberated and transformed by nuclear correlations~\cite{vento-half-skyrmion} to 3 deconfined 1/3-charged quasiquarks. 

That the homogeneous half-skyrmion structure translated into $Gn$EFT works fairly well for massive stars $\gsim 2 M_\odot$~\cite{MR-review} could imply that the $\eta^\prime$ ring if present is metastable so does not intervene  in the density range probed.  Thus the  PC structure could be a possible ``precursor" to a truly deconfined quark phase expected at $n\gsim n_{VM}\gsim 25 n_0$. As noted, this phenomenon could be reflected in the $g_A^L=1$ at density $\sim n_0$ as precursor to $g_A^{DL}=1$ at the dilaton limit fixed point~\cite{WS-invited}..
\section{Further remarks}
I discussed the two potentially important observations re: dense matter. First  the density functional theory \`a la Kohn-Sham -- which is widely exploited in quantum chemistry, condensed matter and nuclear physics -- also captures  FQHE and could be mapped to  Chern-Simons theory, and second,   the physics of the $N_f=1$ and $N_f\geq 2$ baryons could be unified by an EFT  anchored on hidden local symmetry (with the $\rho$ and $\omega$ very familiar in nuclear physics since many decades but with up-to-date unfamiliar charcterstics) combined with scale-symmetry (with the scalar with dilatonic structure), the very same ingredients that figure crucially in going to dense baryonic matter in massive compact stars.  Furthermore  approaching the FQH droplet by deformation by increasing density from the $N_f\geq 2$ baryons at $n\sim n_0$ to  dense compact-star matter seems to reveal uncannily similar fractionalized objects, the former half-pancakes (i.e., pita) and the latter half-skyrmions.  Both involve vanishing condensates with non-vanishing pseudo-scalar meson ($\eta^\prime$, $\pi$) decay constants, both manifested in what appears to be Georgi's ``vector limit" rather than the ``vector manifestation limit". Resolving the dichotomy problem offers a challenge for a new paradigm in nuclear theory for the densest stable stuff in the Universe.

\subsection*{Acknowledgments}
I am grateful for comments from and/or discussions with Hyun Kyu Lee, Yong-Liang Ma, Maciej Nowak and Ismail Zahed.

 \vfil

\end{document}